\documentclass[preprint,superscriptaddress,showpacs,preprintnumbers,amsmath,amssymb]{revtex4}

\usepackage{graphicx}
\usepackage{dcolumn}
\usepackage{bm}
\usepackage{color}

\begin{document}

\title{Instability of superfluid Fermi gases induced by a roton-like density mode in optical lattices}

\author{Yoshihiro Yunomae}
\affiliation{Department of Physics, Waseda University, Okubo,
Shinjuku-ku, Tokyo 169-8555, Japan}
\author{Daisuke Yamamoto}
\affiliation{Department of Physics, Waseda University, Okubo,
Shinjuku-ku, Tokyo 169-8555, Japan}
\author{Ippei Danshita}
\affiliation{Department of Physics, Faculty of Science, Tokyo University
of Science, Kagurazaka, Shinjuku-ku, Tokyo 162-8601, Japan}
\affiliation{Department of Physics, Boston University, Boston, Massachusetts 02215, USA}
\author{Nobuhiko Yokoshi}
\affiliation{Nanotechnology Research Institute, AIST, Tsukuba 305-8568, Japan}
\affiliation{CREST--JST, 4-1-8 Honcho, Saitama 332-0012, Japan}
\author{Shunji Tsuchiya}%
\affiliation{Department of Physics, Keio University, 3-14-1 Hiyoshi, Kohoku-ku, Yokohama 223-8522, Japan}
\affiliation{CREST--JST, 4-1-8 Honcho, Saitama 332-0012, Japan}

\date{\today}

\begin{abstract}
We study the stability of superfluid Fermi gases in deep optical lattices in the BCS--Bose-Einstein condensation (BEC) crossover at zero temperature.
Within the tight-binding attractive Hubbard model, we calculate the
spectrum of the low-energy Anderson-Bogoliubov (AB) 
mode as well as the single-particle excitations in the presence of 
superfluid flow in order to determine the critical velocities.
To obtain the spectrum of the AB mode, we calculate the density response
function in the generalized random-phase approximation applying
the Green's function formalism developed by C\^ot\'e and Griffin to the
Hubbard model.
We find that the spectrum of the AB mode is separated from the particle-hole
continuum having the characteristic rotonlike minimum at short wavelength
due to the strong charge-density-wave fluctuations.
The energy of the rotonlike minimum decreases with increasing the
lattice velocity and it reaches zero at the critical velocity which is
smaller than the pair breaking velocity.
This indicates that the superfluid state is energetically unstable due
to the spontaneous emission of the short-wavelength rotonlike excitations of the
AB mode instead due to pair-breaking. 
We determine the critical velocities as functions of the interaction
strength across the BCS-BEC crossover regime.

\end{abstract}

\maketitle

\section{introduction}

The recent realization of superfluidity in Fermi
gases~\cite{Regal,Bartenstein,Zwierlein,Kinast,Bourdel,CChin} has opened a new research frontier in ultracold atoms \cite{Giorgini}.
A great experimental advantage of this system is the ability in
controlling atomic interactions using a Feshbach resonance~\cite{Ketterle}.
This allows us to access the crossover between the
Bardeen-Cooper-Shrieffer (BCS)-type superfluidity and Bose-Einstein
condensation (BEC) of bound molecules,
which is referred to as the BCS-BEC crossover
\cite{Eagles,Leggett,Nozieres,SadeMelo,Holland,Timmermans,Ohashi,Tamaki}. 
The study of superfluid Fermi gases in the BCS-BEC crossover is expected
to offer new insights into the phenomena of superfluidity and
superconductivity, which can be applied in various fields such
as condensed-matter physics, nuclear physics, and particle physics.

One of the most dramatic features of a superfluid system is the
dissipationless superfluid flow~\cite{Pitaevskii1}.
In particular, critical velocities of superfluid flow have
attracted much interest in various systems such as superfluid $^4$He~\cite{Tilley},
superfluid $^3$He~\cite{Vollhardt}, and atomic Bose-Einstein condensates~\cite{Raman,Onofrio}.
It is well known that the underlying mechanisms for the instability
of dissipationless flow are different in the BCS and BEC regions in a
uniform system. Namely, the instability of BCS-type superfluids is
considered to be dominated by Cooper pair breaking~\cite{Tinkham},
whereas the instability of Bose superfluids is induced by spontaneous
emission of phonon excitations~\cite{Landau}.
It is of interest to study how the mechanism of the instability in
superfluid Fermi gases changes in the BCS-BEC crossover.

Recently, Miller {\it et al}. investigated experimentally the stability
of superfluid flow in Fermi gases in {\it shallow} one-dimensional (1D) optical lattices across the BCS-BEC crossover~\cite{Miller}.
They measured superfluid critical velocities, at which the number
of condensed atoms starts to decrease, by moving the optical lattice
potential through the atomic cloud for different values of 
interatomic interaction and lattice depth~\cite{Miller}.
The measured critical velocities showed a crossover behavior
between the BCS and BEC regimes taking a maximum value at the crossover
regime~\cite{Miller}.
Critical velocities in superfluid Fermi gases in the BCS-BEC
crossover have been also addressed theoretically in several
papers \cite{Combescot,Spuntarelli,Pitaevskii2}.
The observed crossover behavior of the critical velocities has been
predicted in Refs.~\cite{Combescot,Spuntarelli}.
However, most of the theoretical papers are limited within a uniform
system~\cite{Combescot} or a system in the presence of a single potential
barrier~\cite{Spuntarelli}, which cannot be directly compared to
the experiment using optical lattices in Ref.~\cite{Miller}.
In Ref.~\cite{Pitaevskii2}, sound propagation in superfluid Fermi
gases in optical lattices has been studied using the hydrodynamic approximation.
However, microscopic calculation of the critical velocities of
superfluid Fermi gases in optical lattices has not been worked out yet.

In this paper,
we study the stability and critical velocities of superfluid Fermi 
gases in {\it deep} one-dimensional, two-dimensional (2D), and three-dimensional (3D) optical lattices in the BCS-BEC crossover at zero temperature.
We apply the generalized random-phase approximation (GRPA) developed by
C\^ot\'e and Griffin~\cite{Cote} to the attractive tight-binding Hubbard
model in order to calculate the excitation spectra in the presence of a moving
optical lattice.
For the stability of Fermi gases in the BCS-BEC
crossover, two kinds of excitations play crucial roles.
One is the single-particle excitation which arises when Cooper pairs are broken.
The other is the collective density-fluctuation mode, the so-called 
Anderson-Bogoliubov (AB) mode~\cite{Anderson1,Bogoliubov}.
In a uniform system, the single-particle excitation induces the
instability of superfluid flow in the BCS regime, while phonon
excitation of the Bogoliubov mode, which corresponds to the AB mode in
the BCS regime, induces the instability in the BEC
regime~\cite{Combescot,Spuntarelli,Miller}.
We find that in deep 1D, 2D, and 3D optical lattices, the excitation spectrum
of the AB mode has a characteristic rotonlike structure and lies below the
particle-hole continuum due to the strong charge-density-wave (CDW) fluctuation.
The energy of the rotonlike minimum decreases with
increasing the superfluid velocity and it reaches zero before
the particle-hole continuum does, i.e., before pair breaking occurs. 
As a result, in contrast to the uniform case, the instability of superfluid flow in 1D, 2D, and 3D optical lattices is induced by the rotonlike excitations of the AB mode rather than by pair-breakings.
We calculate the critical velocities at which spontaneous emission of
the rotonlike excitations occurs as functions of the interaction
strength in the entire BCS-BEC crossover regime.

This paper is organized as follows.
In Sec.~\ref{formalism}, we present the model and formalism.
We introduce the tight-binding Hubbard model and the Green's
function formalism for the GRPA.
In Sec.~\ref{result}, we present the results for the stability and the
critical velocities of superfluid Fermi gases in 1D, 2D, and 3D optical lattices. 
We calculate the excitation spectra and
determine critical velocities as functions of the attractive interaction.
We summarize our results in Sec.~\ref{conclusion}.

\section{Model and formalism}
\label{formalism}

In this section, we summarize the Green's function formalism applied to
an attractive Hubbard model. This is necessary for the calculation of
response functions in GRPA. The excitation spectra of collective modes
can be obtained as the poles of the response functions. Since our major
interest is in the AB mode, we calculate the density response function
assuming an external field coupled with density.
To discuss the stability of superfluid states, we extend the previous
work for the ground state \cite{Belkhir,Koponen} to the
current-carrying states.

\subsection{Green's function formalism}

We consider two-component atomic superfluid Fermi gases with equal populations
loaded into optical lattices.
We suppose that the optical lattice potential is moving with a constant 
velocity $-\bm v$ in the laboratory frame. If the velocity of the
lattice potential does not exceed the critical velocity, the Fermi gas
remains stable in the laboratory frame due to its superfluidity. This
situation can be described equivalently in the frame fixed with respect
to the lattice potential as a superfluid Fermi gas flowing with a
constant quasimomentum $2m\bm v$, where $m$ is the mass of a fermion.
In the following, we describe the system in the frame fixed with
respect to the lattice potential.
Namely, we assume a time-independent lattice potential and a
supercurrent with the quasimomentum  $2m\bm v$.

We assume that the optical lattice potential is sufficiently deep so that
the tight-binding approximation is valid. 
Thus, the system can be described by a single-band Hubbard model as (we set $\hbar=k_{\rm B}=1$)
\begin{eqnarray}
H=-J\sum_{\langle i,j\rangle,\sigma}\left( c_{i \sigma}^\dagger c_{j\sigma}+{\rm H.c.} \right)
+U\sum_i c_{i\uparrow}^\dagger c_{i\downarrow}^\dagger c_{i\downarrow}c_{i\uparrow}-\mu\sum_{i,\sigma}c^\dagger_{i
 \sigma}c_{i \sigma}\ , 
\label{Hubbard}
\end{eqnarray}
where $c_{j\sigma}$ is the annihilation operator of a fermion on the $j$th site with pseudospin
$\sigma=\uparrow,\downarrow$.
Here, $J$ is the nearest-neighbor hopping energy, $U$ is the
on-site interaction energy, and $\mu$ is the chemical potential. 
We assume an attractive interaction between atoms ($U<0$).

In order to calculate the density response function, we
introduce a {\it fictitious} time-dependent external field $P_j(t)$ which
is coupled with the density.
The Hamiltonian with the external field is given by
\begin{eqnarray}
K(t)&=&H+V(t),\\
V(t)&=&\sum_jP_j(t) n_j,
\end{eqnarray}
where $n_j\equiv \sum_\sigma c^{\dagger}_{j\sigma}c_{j\sigma}$ is the number
operator.
The density response function is obtained by taking a functional
derivative of the single-particle Green's function by the external field. This
will be carried out in Sec.~\ref{responsefn}.

We use the imaginary time Green's function technique \cite{KadanoffBaym}.
The Heisenberg representations of the annihilation and creation operators
in the imaginary time $\tau$ are defined as,
\begin{eqnarray}
c_{j\sigma}(\tau)&=&{\rm exp}\left(\int_0^\tau d\tau^\prime\ K(\tau^\prime)\right)c_{j\sigma}{\rm exp}\left(-\int_0^\tau d\tau^\prime\ K(\tau^\prime)\right),\\
c_{j\sigma}^\dagger(\tau)&=&{\rm exp}\left(\int_0^\tau d\tau^\prime K(\tau^\prime)\right)c_{j\sigma}^\dagger{\rm exp}\left(-\int_0^\tau d\tau^\prime\ K(\tau^\prime)\right).
\end{eqnarray}
We introduce the normal and anomalous single-particle Green's functions,
respectively, as \cite{Gorkov}
\begin{eqnarray}
G_{ij,\sigma}(\tau,\tau^\prime)&=&-\langle T(c_{i\sigma}(\tau)c^{\dagger}_{j\sigma}(\tau^\prime))\rangle, \\
F_{ij}(\tau,\tau^\prime)&=&-\langle T(c_{i\uparrow}(\tau)c_{j\downarrow}(\tau^\prime))\rangle,
\end{eqnarray}
where $T(\cdots)$ represents the time-ordering operator
with respect to $\tau$. 
Using the Nambu representation \cite{Nambu} with
\begin{eqnarray}
\Psi_j(\tau)=
\left(
\begin{array}{c}
c_{j\uparrow}(\tau) \\
c^{\dagger}_{j\downarrow}(\tau)\\
\end{array}
\right);\ \ 
\Psi_j^\dagger(\tau)=\left(c_{j\uparrow}^\dagger(\tau), c_{j\downarrow}(\tau)\right),
\label{Nambu}
\end{eqnarray}
the single-particle Green's function can be written in the matrix form as
\begin{eqnarray}
\hat{G}_{ij}(\tau, \tau^\prime)\!&\equiv&\!-\langle T(\Psi_i(\tau)\Psi_j^{\dagger}(\tau^\prime))\rangle \nonumber \\
\!&=&\!\left(
\begin{array}{cc}
G_{ij,\uparrow}(\tau,\tau^\prime) & F_{ij}(\tau,\tau^\prime) \\
F^{\ast}_{ij}(\tau,\tau^\prime)  & -G_{ji,\downarrow}(\tau^\prime,\tau) \\
\end{array}
\right).
\label{mxGreen}
\end{eqnarray}
We note that in the absence of the external field, the Green's
function at equal sites and imaginary times $\hat G_{jj}(\tau,\tau)$ is
given by
\begin{eqnarray}
\hat G_{jj}(\tau,\tau)&\equiv&\lim_{\tau^\prime\to\tau+0}\hat G_{jj}(\tau,\tau^\prime)\\
&=&
\left(
\begin{array}{cc}
\langle n_{j\uparrow}\rangle & \langle m_j\rangle \\
\langle m_j\rangle^\ast & 1-\langle n_{j\downarrow}\rangle
\end{array}
\right),
\label{equalG}
\end{eqnarray}
where $n_{j\sigma}\equiv c_{j\sigma}^\dagger c_{j\sigma}$ and $m_j\equiv
c_{j\downarrow}c_{j\uparrow}$ is the pair annihilation operator which is
related to the wave function of Cooper pairs, as we discuss below.

From the equations of motion for $c_{j\sigma}(\tau)$ and
$c_{j\sigma}^\dagger(\tau)$, we obtain the equation for the matrix Green's
function in Eq.~(\ref{mxGreen}), as
\begin{eqnarray}
&&\left(-\frac{\partial}{\partial\tau}+\mu \hat\sigma_3\right)
\hat G_{ij}(\tau,\tau^\prime)
+2J\sum_{\langle l,m
\rangle}\delta_{i,l}\hat\sigma_3\hat{G}_{mj}(\tau,\tau^\prime)\nonumber \\
&&=\delta_{i,j}\delta(\tau-\tau^\prime)-U\hat\sigma_3
 \langle
 T(n_i(\tau)\Psi_i(\tau)\Psi^\dagger_j(\tau^\prime))
 \rangle +P_i(\tau)\hat{\sigma}_3 \hat G_{ij}(\tau,\tau^\prime),
\label{motionG}
\end{eqnarray}
where $\hat{\sigma}_3$ is the Pauli matrix
\begin{eqnarray}
\hat\sigma_3=
\left(
\begin{array}{cc}
1 & 0  \\
0 & -1 \\
\end{array}
\right).
\end{eqnarray}
The non-interacting Green's function $\hat G^0_{ij}(\tau,\tau^\prime)$ satisfies
\begin{eqnarray}
\left(
 -\frac{\partial}{\partial\tau}+\mu\hat\sigma_3\right)\hat G^0_{ij}(\tau,\tau^\prime)
+2J\sum_{\langle l,m 
\rangle}\delta_{i,l}\hat\sigma_3\hat G_{mj}(\tau,\tau^\prime)=\delta_{i,j}\delta(\tau-\tau^\prime). 
\label{G0}
\end{eqnarray}
From Eqs.~(\ref{motionG}) and (\ref{G0}), the Green's function
satisfies the Dyson equation
\begin{eqnarray}
\hat G_{ij}(\tau,\tau^\prime)&=&\hat G^0_{ij}(\tau,\tau^\prime)\nonumber\\
&+&\sum_{l,m}\int_0^{\beta} d\tau_1\int_0^{\beta} d\tau_2\ \hat
 G^0_{il}(\tau,\tau_1)\hat \Sigma_{lm}(\tau_1,\tau_2)\hat G_{mj}(\tau_2,\tau^\prime) \nonumber \\
&+&\sum_l\int_0^{\beta} d\tau_1\ \hat G^0_{il}(\tau,\tau_1)P_l(\tau_1)
 \hat{\sigma}_3\hat G_{lj}(\tau_1,\tau^\prime), \label{Dyson}
\end{eqnarray}
where $\beta=1/T$ and $T$ is the temperature. 
In Eq.~(\ref{Dyson}), the self-energy $\hat\Sigma_{ij}(\tau,\tau^\prime)$ is given by
\begin{eqnarray}
\hat\Sigma_{ij}(\tau, \tau^\prime)
=-U\hat{\sigma}_3\sum_l\int_0^\beta d\tau_1\  \langle T (n_i(\tau)
\Psi_i(\tau) \Psi^\dagger_l (\tau_1)) [\hat G_{lj}(\tau_1,\tau^\prime)]^{-1}.
\label{self}
\end{eqnarray}
Here, we introduced the inverse matrix Green's function $[\hat G_{ij}(\tau,\tau^\prime)]^{-1}$,
which satisfies
\begin{eqnarray}
\sum_l\int_{0}^{\beta} d\tau_1\ \hat G_{il}(\tau,\tau_1)
(\hat G_{lj}(\tau_1,\tau^\prime))^{-1}=\delta_{i,j}\delta(\tau-\tau^\prime).
\label{G-1}
\end{eqnarray}
Using Eq.~(\ref{G-1}), Eq.~(\ref{Dyson}) can be simplified as
\begin{eqnarray}
[\hat G_{ij}(\tau,\tau^\prime)]^{-1}=[\hat G^0_{ij}(\tau,\tau^\prime)]^{-1}-\hat\Sigma_{ij}(\tau,\tau^\prime)-P_i(\tau)\hat{\sigma}_3\delta_{i,j}\delta(\tau-\tau^\prime).
\label{Dyson-1}
\end{eqnarray}

To calculate the self-energy in Eq.~(\ref{self}), we use the
Hartree-Fock-Gor'kov (HFG) approximation \cite{Cote,Gorkov}
\begin{equation}
\langle T(n_l(\tau_2)\Psi_i(\tau)\Psi^\dagger_j(\tau_1)) \rangle
\simeq -\langle n_l(\tau_2)\rangle\hat{G}_{ij}(\tau,\tau_1) 
+\hat{G}_{il}(\tau,\tau_2)\hat{\sigma}_3\hat{G}_{lj}(\tau_2,\tau_1).
\label{HFG}
\end{equation}
Thus, the self-energy in the HFG approximation is given by
\begin{eqnarray}
\hat{\Sigma}_{ij}(\tau,\tau^\prime) &\simeq& \hat{\Sigma}^{\rm HFG}_{ij}(\tau,\tau^\prime) \nonumber \\
&=&U[ \langle n_i(\tau) \rangle \hat{\sigma}_3-\hat{\sigma}_3\hat{G}_{ii}(\tau,\tau)\hat\sigma_3]\delta_{i,j}\delta(\tau-\tau^\prime).
\end{eqnarray}

In the presence of supercurrent with velocity $\bm v$, Cooper pairs
are Bose-condensed into the state with the center-of-mass quasimomentum
$\bm q=2m\bm v$.
Since the anomalous Green's function $F_{jj}(\tau,\tau)$ can be regarded
as the wave function of Cooper pairs~\cite{Gorkov}, it can be written as
\begin{eqnarray}
F_{jj}(\tau,\tau)&=&-\langle m_j\rangle=\frac{\Delta_{\bm v}}{|U|}{\rm exp}(2im{\bm v}\cdot{\bm r}_j),
\label{wf}
\end{eqnarray}
where $\Delta_{\bm v}$ is the superfluid gap and $\bm r_j$ is the location of the $j$th site.
The exponential factor on the right-hand side of Eq.~(\ref{wf})
describes the supercurrent with quasimomentum $2m\bm v$.

In the presence of the supercurrent, the normal and anomalous Green's
functions can be written as
\begin{eqnarray}
G_{ij,\sigma}(\tau,\tau^\prime)&=&{\rm exp}(im\bm v\cdot \bm r_{ij})\tilde G_{ij,\sigma}(\tau,\tau^\prime),\\
F_{ij}(\tau,\tau^\prime)&=&{\rm exp}(2im\bm v\cdot \bm R_{ij})\tilde
 F_{ij}(\tau,\tau^\prime),\label{Ftilde}
\end{eqnarray}
respectively, where $\bm r_{ij}\equiv\bm r_i-\bm r_j$ is the relative coordinate and $\bm
R_{ij}\equiv(\bm r_i+\bm r_j)/2$ is the center-of-mass coordinate of the
Cooper pair.
Here, $\tilde G_{ij,\sigma}(\tau,\tau^\prime)$ and $\tilde F_{ij}(\tau,\tau^\prime)$ are functions of $\bm r_{ij}$.

To eliminate the phase factors associated with the supercurrent, it is convenient to introduce an operator
$\tilde\Psi_j(\tau)$ and a matrix Green's function $\hat{\tilde
G}_{ij}(\tau,\tau^\prime)$ as,
\begin{eqnarray}
\tilde\Psi_j(\tau)&=&
\left(
\begin{array}{l}
c_{j\uparrow}(\tau){\rm exp}(-im\bm v\cdot \bm r_j)\\
c_{j\downarrow}^\dagger(\tau) {\rm exp}(im\bm v\cdot \bm r_j)
\end{array}
\right)=\hat\gamma_j\Psi_j(\tau),\\
\hat{\tilde G}_{ij}(\tau, \tau^\prime)&=&-\langle
 T(\tilde\Psi_i(\tau)\tilde\Psi_j^\dagger(\tau^\prime))\rangle\nonumber\\
&=&\hat\gamma_i\hat{G}_{ij}(\tau,\tau^\prime)\hat\gamma^\ast_j\nonumber\\
&=&
\left(
\begin{array}{cc}
\tilde G_{ij,\uparrow}(\tau,\tau^\prime) & \tilde F_{ij}(\tau,\tau^\prime) \\
\tilde F_{ij}(\tau,\tau^\prime) & -\tilde G_{ji,\downarrow}(\tau^\prime,\tau) \\
\end{array}
\right),
\label{Green}
\end{eqnarray}
where the matrix $\hat\gamma_j$ for the unitary transformation between $\hat G$
and $\hat{\tilde G}$ is given by
\begin{equation}
\hat{\gamma}_j=\left(
\begin{array}{cc}
{\rm exp}(-im{\bm v}\cdot {\bm r}_j) & 0  \\
0 & {\rm exp}(i m{\bm v}\cdot{\bm r}_j) \\
\end{array}
\right).
\end{equation}
Using Eq.~(\ref{equalG}) and (\ref{Green}), in the absence of the external field,
$\hat{\tilde G}_{jj}(\tau,\tau)$ reduces to
\begin{equation}
\hat{\tilde G}_{jj}(\tau,\tau)=
\left(
\begin{array}{cc}
\langle n_{j\uparrow}\rangle & \Delta_{\bm v}/|U|  \\
\Delta^\ast_{\bm v}/|U| & 1-\langle n_{j\downarrow}\rangle \\
\end{array}
\right).
\end{equation}
In the following sections, we derive equations for $\hat{\tilde G}$.

\subsection{Equilibrium Green's function}

In this section, we calculate the equilibrium Green's function in the
absence of the external field $P_j(t)$ within the HFG approximation
introduced in Eq.~(\ref{HFG}).
Since $\hat{\tilde G}_{ij}(\tau,\tau^\prime)$ in Eq.~(\ref{Green}) is a function of $\bm r_{ij}$, we define
the Fourier transform of the Green's function as
\begin{eqnarray}
\hat{\tilde{G}}_{ij}(\tau,\tau^\prime)=\frac{1}{\beta M}\sum_{\bm k,\omega_n} {\rm exp}[i\bm k\cdot\bm r_{ij}-i\omega_n(\tau-\tau^\prime)]\hat{\tilde{G}}_{\bm k}(i\omega_n), \label{fourierG}
\end{eqnarray}
where $M$ is the number of lattice sites and
$\omega_n\equiv(2n+1)\pi/\beta$ is the Fermi Matsubara frequency.
We note that $\hat G_{ij}(\tau,\tau^\prime)$ cannot be expanded as Eq.~(\ref{fourierG}) because
the phase factor in Eq.~(\ref{Ftilde}) describing the supercurrent depends on
the center-of-mass coordinate $\bm R_{ij}$.

From Eq.~(\ref{Dyson}), we obtain the Dyson equation in Fourier space in the
absence of the external field as
\begin{eqnarray}
\hat{\tilde{G}}_{\bm k}(i\omega_n)=\hat{\tilde{G}}^0_{\bm k}(i\omega_n)+\hat{\tilde{G}}^0_{\bm k}(i\omega_n)\hat{\tilde{\Sigma}}_{\bm k}(i\omega_n)\hat{\tilde{G}}_{\bm k}(i\omega_n). \label{Dysonk}
\end{eqnarray}
Here, $\hat{\tilde \Sigma}_{\bm k}(i\omega_n)$ is the Fourier transform
of $\hat{\tilde \Sigma}_{ij}(\tau,\tau^\prime)\equiv \hat\gamma_i\hat\Sigma_{ij}(\tau,\tau^\prime)\hat\gamma_j^\ast$.
In Eq.~(\ref{Dysonk}), the unperturbed Green's function
$\hat{\tilde{G}}^0_{\bm k}(i\omega_n)$ is given by
\begin{eqnarray}
\hat{\tilde{G}}^0_{\bm k}(i\omega_n)=\left(
\begin{array}{cc}
{\displaystyle \frac{1}{i\omega_n-\xi_{\bm k+m\bm v}}}&0 \\
0&{\displaystyle \frac{1}{i\omega_n+\xi_{\bm k-m\bm v}}} \\
\end{array} 
\right),
\end{eqnarray}
where $\xi_{\bm k}=2J\sum_{\nu}(1-\cos k_\nu d)-\mu$ is the kinetic
energy, $\nu$ is the index for spatial
dimension, and $d$ is the lattice constant.
From Eq.~(\ref{HFG}), we obtain the self-energy in Eq.~(\ref{Dysonk}) as
\begin{eqnarray}
\hat{\tilde{\Sigma}}_{\bm k}(i\omega_n)=\left(
\begin{array}{cc}
0&\Delta_{\bm v} \\
\Delta^{\ast}_{\bm v}&0 \\
\end{array} 
\right).
\label{selfk}
\end{eqnarray}
In deriving Eq.~(\ref{selfk}), we shifted the chemical potential by the
Hartree-Fock energy $nU/2$, where $n$ is the average number of atoms per
site.

\begin{figure}
\centerline{\includegraphics{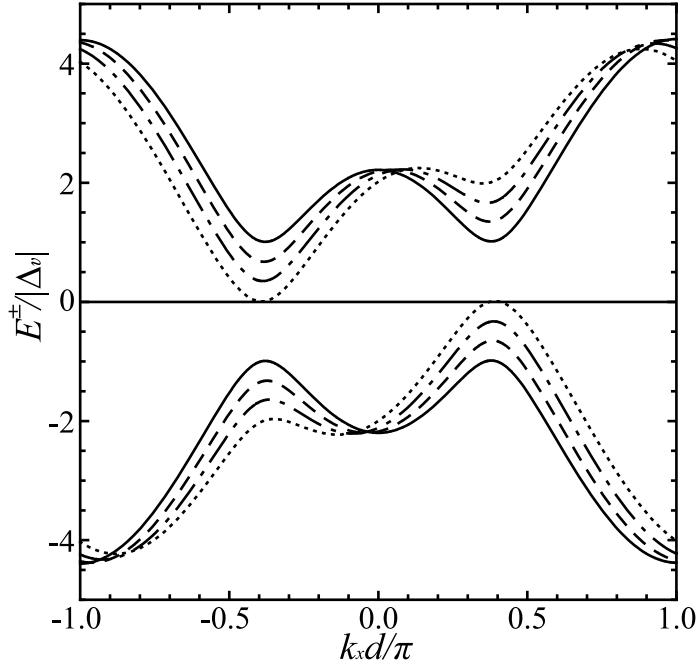}}
\caption{Single-particle excitation spectra $E^+_{\bm k}$ (upper curves) and
 $E^-_{\bm k}$ (lower curves) in 3D optical lattices when the superfluid velocity $|{\bm v}|$ is 0 (solid line), $0.2/md$ (dashed line), $0.4/md$ (dash-dotted line), and $|\bm v_{\rm pb}|$ (dotted line). Here, the superfluid flows along the $(\pi,\pi,\pi)$ direction and $|\bm v_{\rm pb}|=0.628/md$. We set $U/J=-6.0$, $n=0.5$, and $k_x=k_y=k_z$.}
\label{spex}
\end{figure}

By solving Eq.~(\ref{Dysonk}) with the self-energy in Eq.~(\ref{selfk}),
we obtain the single-particle Green's function as
\begin{equation}
\hat{\tilde{G}}_{\bm k}(i\omega_n)=
\frac{\hat A_{\bm k}}{i\omega_n-E^{+}_{\bm k}}+\frac{\hat B_{\bm
k}}{i\omega_n-E^{-}_{\bm k}}, 
\label{Single-Green}
\end{equation}
where
\begin{eqnarray}
\hat A_{\bm k}&=&
\left(
\begin{array}{cc}
u_{\bm k}^2 & u_{\bm k}v^{\ast}_{\bm k} \\
u_{\bm k}v_{\bm k} & |v_{\bm k}|^2 \\
\end{array} 
\right),\\
\hat B_{\bm k}&=&
\left(
\begin{array}{cc}
|v_{\bm k}|^2&-u_{\bm k}v^{\ast}_{\bm k} \\
-u_{\bm k}v_{\bm k}&u_{\bm k}^2 \\
\end{array} 
\right),\\
u_{\bm k}^2&=&\frac{1}{2}\left(1+\frac{\bar{\xi}_{\bm k}}{\mathcal{E}_{\bm k}} \right),\\
|v_{\bm k}|^2&=&\frac{1}{2}\left(1-\frac{\bar{\xi}_{\bm
			    k}}{\mathcal{E}_{\bm k}} \right), \\
u_{\bm k}v^{\ast}_{\bm k}&=&\frac{\Delta_{\bm v}}{2 \mathcal{E}_{\bm k}}.
\end{eqnarray}
Here, the single-particle excitation energy is given by
\begin{equation}
E^{\pm}_{\bm k}=\eta_{\bm k}\pm\mathcal E_{\bm k},
\label{spspectrum}
\end{equation}
where $\mathcal{E}_{\bm k}=\sqrt{\bar{\xi}^2_{\bm k}+|\Delta_{\bm v}|^2}$,
$\bar{\xi}_{\bm k}=(\xi_{\bm k+m\bm v}+\xi_{\bm k-m\bm
v})/2$, and $\eta_{\bm k}=(\xi_{\bm k+m\bm v}-\xi_{\bm k-m\bm v})/2$.
Equation~(\ref{spspectrum}) explicitly shows that $E^{\pm}_{\bm k}$ depend on the superfluid velocity $\bm v$.
The single-particle excitation spectrum in Eq.~(\ref{spspectrum}) is
shown in Fig.~\ref{spex} for different superfluid velocities. 
In Fig.~\ref{spex}, the energy gap becomes smaller as
$|\bm v|$ increases. When the energy gap reaches zero, pair breaking occurs,
i.e., ${\bm v}={\bm v}_{\rm pb}$, where ${\bm v}_{\rm pb}$ is the pair-breaking velocity \cite{Rodriguez}.

We determine the superfluid gap $\Delta_{\bm v}$ and the chemical
potential $\mu$ by solving self-consistently the number equation 
\begin{eqnarray}
n=\frac{2}{M} \sum_{\bm k} \left[ (u_{\bm k}^2-|v_{\bm k}|^2)
			    f(E^{+}_{\bm k}) +|v_{\bm k}|^2\right]
\label{numbereq}
\end{eqnarray}
and the gap equation
\begin{eqnarray}
\Delta_{\bm v}=-\frac{U}{M} \sum_{\bm k}u_{\bm k}v_{\bm k}^\ast [1-2f(E^+_{\bm k})]. \label{gapeq}
\end{eqnarray}
Equations~(\ref{numbereq}) and (\ref{gapeq}) are obtained from
the diagonal and off-diagonal elements of the Green's function in Eq.~(\ref{Single-Green}).
Here, $f(\varepsilon)\equiv 1/[\exp(\beta \varepsilon)+1]$ is the Fermi distribution function.
This scheme of solving Eqs.~(\ref{numbereq}) and (\ref{gapeq}) self-consistently
interpolates the weak-coupling BCS limit and strong-coupling 
BEC limit at low temperature when the fluctuation effect due to pairs
with finite center of mass momenta can be neglected~\cite{Leggett}. 
Throughout the work including the calculation of $\Delta_{\bm v}$ and $\mu$, we assume $T=0$.

\begin{figure}
\centerline{\includegraphics{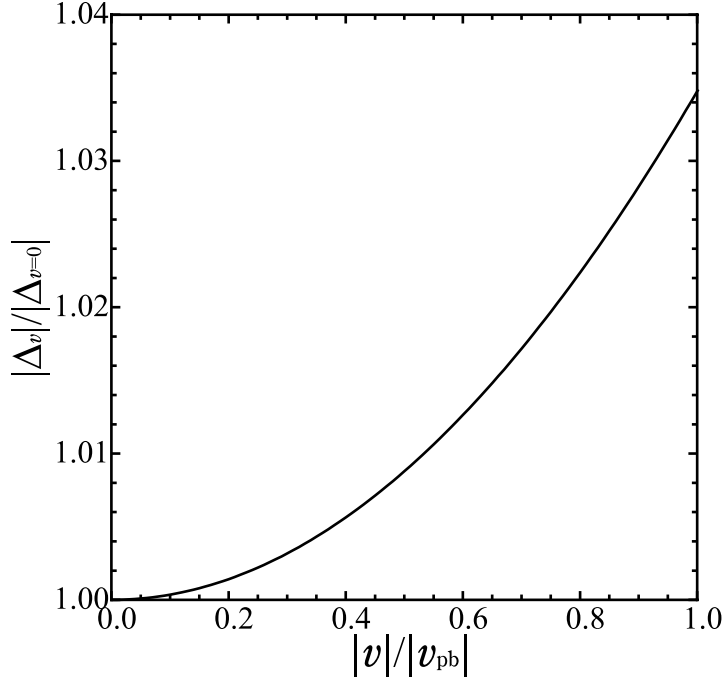}}
\caption{Superfluid gap $\Delta_{\bm v}$ as a function of $|\bm v|$ in 3D optical lattices, when the superfluid flows along the $(\pi,\pi,\pi)$ direction. Here,  $|\Delta_{{\bm v}=0}|=1.92J$ and $|\bm v_{\rm pb}|=0.628/md$. We set $U/J=-6.0$ and $n=0.5$.}
\label{gap}
\end{figure}

In Eq. (40), $\Delta_{\bm v}$ depends on superfluid velocity $\bm v$ via $u_{\bm k}$, $v_{\bm k}$, and $E_{\bm k}^+$. To explicitly show this, we plot $\Delta_{\bm v}$ in 3D at $T=0$ as a function of $|{\bm v}|(\leq |{\bm v}_{\rm pb}|)$ in Fig. \ref{gap}. Note that when $E_{\bm k}^+=0$ at $|{\bm v}|=|{\bm v}_{\rm pb}|$, the superfluid gap does not vanish ($\Delta_{\bm v}\neq 0$), but the superfluid state is destabilized due to pair breaking.

\subsection{Response function}
\label{responsefn}

In this section, we calculate the density response function.
The density response function can be derived by taking a functional
derivative of the density by the external field as
\begin{equation}
\chi_{ij}(\tau,\tau^\prime)=\frac{\delta\langle n_{i}(\tau)\rangle}{\delta P_j(\tau^\prime)}. \label{chi}
\end{equation}
Here, we introduce the three-point correlation function
$\hat L_{ijl}(\tau,\tau^\prime,\tau_1)$ as
\begin{equation}
\hat L_{ijl}(\tau,\tau^\prime,\tau_1)\equiv \frac{\delta
 \hat{\bar{G}}_{ij}(\tau,\tau^\prime)}{\delta P_l(\tau_1)},
\label{3point}
\end{equation}
where $\hat{\bar{G}}_{ij}(\tau,\tau^\prime)\equiv\sigma_3\hat{\tilde{G}}_{ij}(\tau,\tau^\prime)$.
Using Eq.~(\ref{3point}), the density response function can be written as
\begin{eqnarray}
\chi_{ij}(\tau,\tau^\prime)&=&-\langle T(\delta n_i(\tau)\delta n_j(\tau^\prime))\rangle\nonumber\\
&=&L^{11}_{ij}(\tau,\tau^\prime)+L^{22}_{ij}(\tau,\tau^\prime),\label{chi2}
\end{eqnarray}
where $\delta n_i(\tau)\equiv n_i(\tau)-\langle n_i(\tau)\rangle$ is the
density fluctuation operator and 
\begin{equation}
\hat{L}_{ij}(\tau,\tau^\prime)\equiv \lim_{\tau_1 \to \tau^{+}} \hat{L}_{iij}(\tau,\tau_1,\tau^\prime).
\end{equation}
In deriving Eq.~(\ref{chi2}), we used the functional differentiation of
the single-particle Green's function by the external field as
\begin{equation}
\frac{\delta \hat{\tilde{G}}_{ij}(\tau,\tau^\prime)}{\delta
 P_{l}(\tau_1)}=\langle T(\tilde\Psi_i(\tau) \tilde\Psi^\dagger_j(\tau^\prime)
 n_{l}(\tau_1) )\rangle+\hat{\tilde{G}}_{ij}(\tau,\tau^\prime)\langle
 n_l(\tau_1) \rangle. 
\label{difG}
\end{equation}
We note that Eq.~(\ref{difG}) is valid within the linear-response regime.

We derive the equation of motion for the three-point correlation
function $\hat L$. 
Differentiating Eq.~(\ref{G-1}) with respect to the external field, we obtain
\begin{eqnarray}
\frac{\delta \hat{\tilde{G}}_{ij}(\tau, \tau^\prime)}{\delta P_l(\tau_1)}
&=&\sum_{m,n}\int_0^\beta d\tau_2 \int_{0}^{\beta}\ d\tau_3\ \hat{\tilde{G}}_{im}(\tau,\tau_2)\frac{\delta \hat{\tilde{\Sigma}}_{mn}(\tau_2,\tau_3)}{\delta P_l(\tau_1)}\hat{\tilde{G}}_{nj}(\tau_3,\tau^\prime)\nonumber\\
&+&\sum_{m}\int_0^\beta d\tau_2\ \hat{\tilde{G}}_{im}(\tau,\tau_2)\frac{\delta P_{m}(\tau_2)}{\delta P_l(\tau_1)}\hat{\sigma}_3\hat{\tilde{G}}_{mj}(\tau_2,\tau^\prime).
\end{eqnarray}
Using the HFG approximation in Eq.~(\ref{HFG}), the three-point correlation
function satisfies
\begin{eqnarray}
\hat L_{ijl}(\tau,\tau^\prime,\tau_1)&=&\hat L^0_{ijl}(\tau,\tau^\prime,\tau_1)+U\sum_m\int_{0}^{\beta} d\tau_2\ \hat{\bar{G}}_{im}(\tau,\tau_2)\hat{\bar{G}}_{mj}(\tau_2,\tau^\prime)\chi_{ml}(\tau_2,\tau_1) \nonumber \\
&-&U\sum_m\int_{0}^{\beta}
 d\tau_2\ \hat{\bar{G}}_{im}(\tau,\tau_2)\hat{L}_{ml}(\tau_2,\tau^\prime)\hat{\bar{G}}_{mj}(\tau_2,\tau_1),
\label{Lij} 
\end{eqnarray}
where the lowest-order correlation function is given by
$\hat{L}^0_{ijl}(\tau,\tau^\prime,\tau_1)=\hat{\bar{G}}_{il}(\tau,\tau_1)\hat{\bar{G}}_{lj}(\tau_1,\tau^\prime)$.
Thus, the density response function can be obtained by solving
Eq.~(\ref{Lij}) which is referred to as the GRPA equation \cite{Cote}.

To see the diagrammatic structure of Eq.~(\ref{Lij}) more clearly, it is
useful to rewrite Eq.~(\ref{Lij}) in terms of the irreducible correlation 
function 
\begin{equation}
\hat{\bar{L}}_{ijl}(\tau,\tau^\prime,\tau_1)=\hat{L}^0_{ijl}(\tau,\tau^\prime,\tau_1)
-U\sum_m \int_0^\beta d\tau_2\ \hat{\bar{G}}_{im}(\tau,\tau_2)\hat{\bar{L}}_{ml}(\tau_2,\tau_1)\hat{\bar{G}}_{mj}(\tau_2,\tau^\prime). 
\label{ladder} 
\end{equation}
Using Eq.~(\ref{ladder}), Eq.~(\ref{Lij}) reduces to
\begin{equation}
\hat{L}_{ijl}(\tau,\tau^\prime,\tau_1)=\hat{\bar{L}}_{ijl}(\tau,\tau^\prime,\tau_1)+U\sum_m
 \int_0^\beta d\tau_2\
 \hat{\bar{L}}_{ijm}(\tau,\tau^\prime,\tau_2)\chi_{ml}(\tau_2,\tau_1).
\label{bubble}
\end{equation}
It is clear from Eq.~(\ref{ladder}) that $\hat{\bar L}$ includes the
ladder diagrams. On the other hand, Eq.~(\ref{bubble}) includes the
bubble diagrams which lead to the random-phase approximation (RPA)~\cite{Cote}.
In a homogeneous system, the contribution from the bubble diagrams for an
attractive interaction can be neglected \cite{Cote}. 
However, in our lattice system, since the bubble diagrams
induce the instability due to the CDW fluctuation, it is crucial for
the analysis of the stability of the system to keep the bubble diagrams
in Eq.~(\ref{bubble}). We compare the excitation spectra with and
without the contribution from the bubble diagrams and give a detailed
discussion of the effects of the CDW fluctuations on the stability of
the system in Sec.~\ref{result}.

We solve Eqs. (\ref{ladder}) and (\ref{bubble}) to
calculate the density response function in Eq.~(\ref{chi}).
We define the Fourier transform of $\hat{L}_{ij}(\tau,\tau^\prime)$ as
\begin{eqnarray}
\hat{L}_{ij}(\tau,\tau^\prime)=\frac{1}{\beta M}\sum_{\bm q,\Omega_n}
 {\rm exp}[i\bm q\cdot\bm r_{ij}-i\Omega_n (\tau-\tau^\prime)]\hat L_{\bm q}(i\Omega_n),
 \label{fourierL}
\end{eqnarray}
where $\Omega_n\equiv 2n\pi/\beta$ is the Bose Matsubara frequency.
From Eq.~(\ref{chi2}), the Fourier component of the density response function is given
by $\chi_{\bm q}(i\Omega_n)=L_{\bm q}^{11}(i\Omega_n)+L_{\bm q}^{22}(i\Omega_n)$.

The Fourier transforms of Eqs. (\ref{ladder}) and
(\ref{bubble}) are represented as
\begin{eqnarray}
\hat{\bar{L}}_{\bm q}(i\Omega_n)= \hat{L}^0_{\bm
 q}(i\Omega_n)-\frac{U}{\beta M}\sum_{\bm k,\omega_n} \hat{\bar{G}}_{\bm
 k}(i\omega_n) \hat{\bar{L}}_{\bm q}(i\Omega_n) \hat{\bar{G}}_{\bm k-\bm
 q}(i\omega_n-i\Omega_n) \label{ladderk}
\end{eqnarray}
and
\begin{eqnarray}
\hat L_{\bm q}(i\Omega_n)= \hat{\bar{L}}_{\bm q}(i\Omega_n)+U \hat{\bar{L}}_{\bm q}(i\Omega_n)\chi_{\bm q}(i\Omega_n), \label{bubblek}
\end{eqnarray}
respectively, where
\begin{eqnarray}
&&\hat{L}^{0}_{\bm q}(i\Omega_n)=\frac{1}{\beta M}\sum_{\bm k,\omega_n}\hat{\bar{G}}_{\bm k}(i\omega_n)\hat{\bar{G}}_{\bm k-\bm q}(i\omega_n-i\Omega_n).
\end{eqnarray}
In order to rewrite Eqs.~(\ref{ladderk}) and (\ref{bubblek}) in simpler
forms, we define a column vector $\mathcal{L}_{\bm q}(i\Omega_n)$ as
\begin{eqnarray}
\mathcal{L}_{\bm q}(i\Omega_n)\equiv\left(
\begin{array}{c}
L^{11} \\
L^{12} \\
L^{21} \\
L^{22} \\
\end{array} 
\right).
\end{eqnarray}
Here, we have used the notation $L^{\mu\nu}\equiv L_{\bm q}^{\mu\nu}(i\Omega_n)$.
The same notation is adapted for $\bar{\mathcal{L}}_{\bm q}(i\Omega_n)$ and
$\mathcal{L}_{\bm q}^0(i\Omega_n)$. 
In addition, we define a $4\times 4$ matrix $\hat{\mathcal D}$ as
\begin{equation}
\hat{\mathcal{D}}_{\bm q}(i\Omega_n)
\equiv\left(
\begin{array}{cccc}
D^{1111} & D^{1121} & D^{1211} & D^{1221} \\
D^{1112} & D^{1122} & D^{1212} & D^{1222} \\
D^{2111} & D^{2121} & D^{2211} & D^{2221} \\
D^{2112} & D^{2122} & D^{2212} & D^{2222} \\
\end{array} 
\right),
\label{matrixD}
\end{equation}
where
\begin{equation}
D_{\bm q}^{\mu\nu\rho\lambda}(i\Omega_n)\equiv\frac{1}{\beta M}\sum_{\bm k,\omega_n}\bar{G}^{\mu\nu}_{\bm k+\bm q}(i\omega_n+i\Omega_n)\bar{G}^{\rho\lambda}_{\bm k}(i\omega_n).\label{A04}
\end{equation}
From Eq.~(\ref{A04}), $\hat D$ describes a single bubble diagram of a particle-hole excitation.
Thus, Eqs.~(\ref{ladderk}) and (\ref{bubblek}) can be written in the
matrix forms as
\begin{eqnarray}
\bar{\mathcal L}_{\bm q}(i\Omega_n)&=& {\mathcal L}^0_{\bm
 q}(i\Omega_n)-U\hat{\mathcal D}_{\bm q}(i\Omega_n) \bar{\mathcal L}_{\bm q}(i\Omega_n),\label{ladder4} \\
{\mathcal L}_{\bm q}(i\Omega_n)&=&\bar{\mathcal L}_{\bm q}(i\Omega_n)+U
 \bar{\mathcal L}_{\bm q}(i\Omega_n)\chi_{\bm q}(i\Omega_n), \label{bubble4}
\end{eqnarray}
respectively.
Solving Eq.~(\ref{ladder4}), we obtain
\begin{eqnarray}
\bar{\mathcal L}_{\bm q}(i\Omega_n)=[ \hat{I}+U\hat{\mathcal
				     D}_{\bm q}(i\Omega_n)
				    ]^{-1}{\mathcal L}^0_{\bm q}(i\Omega_n).\label{lbar}
\end{eqnarray}
Here, $\hat{I}$ is the $4\times 4$ unit matrix.
After the analytic continuation
$i\Omega_n\rightarrow\omega+i\delta$ (we take the limit $\delta\rightarrow +0$ after the calculation), we obtain the density response function as 
\begin{eqnarray}
\chi_{\bm q}(\omega)=\frac{[\mathcal{\bar{L}}_{\bm q}(\omega)]_1+[\mathcal{\bar{L}}_{\bm q}(\omega)]_4}{1-U\{[\mathcal{\bar{L}}_{\bm q}(\omega)]_1+[\mathcal{\bar{L}}_{\bm q}(\omega)]_4\}}.\label{chiGRPA}
\end{eqnarray}
The excitation spectrum of the AB mode is obtained from the pole of
Eq.~(\ref{chiGRPA}).

If we only take into account the ladder diagrams, $\chi_{\bm q}(\omega)$ reduces to
\begin{eqnarray}
\chi^{\rm L}_{\bm q}(\omega)=[\mathcal{\bar{L}}_{\bm q}(\omega)]_1+[\mathcal{\bar{L}}_{\bm q}(\omega)]_4. \label{chiladder}
\end{eqnarray}
Equation~(\ref{chiladder}) shows that the pole of $\chi^{\rm L}(\omega)$
coincides with that of $\bar{\mathcal L}_{\bm q}(\omega)$ which is
obtained from the condition $|\hat I+U\hat{\mathcal D}_{\bm q}(\omega)|=0$ in Eq.~(\ref{lbar}).

We calculate the single bubble diagram $\hat{\mathcal
D}_{\bm q}(\omega)$ in Eq.~(\ref{A04}).
Substituting Eq.~(\ref{Single-Green}) into Eq.~(\ref{A04}), we obtain
\begin{eqnarray}
D_{\bm q}^{\mu\nu\rho\lambda}(\omega)&=&\frac{1}{M}\sum_{\bm k}
 \left[\bar{A}_{\bm k+\bm q}^{\mu\nu}\bar{A}_{\bm
  k}^{\rho\lambda}\frac{f(E^+_{\bm k})-f(E^+_{\bm k+\bm
  q})}{\omega+E^+_{\bm k} -E^+_{\bm k+\bm q}+i\delta}-\bar{B}_{\bm k-\bm
  q}^{\mu\nu}\bar{B}_{\bm k}^{\rho\lambda}\frac{f(E^+_{\bm k})-f(E^+_{\bm
  k-\bm q})}{\omega-E^+_{\bm k} +E^+_{\bm k-\bm q}+i\delta}\right.\nonumber\\
&&\left.+\bar{A}_{\bm k-\bm q}^{\mu\nu}\bar{B}_{\bm k}^{\rho\lambda}\frac{1-f(E^+_{\bm k})-f(E^+_{-\bm k+\bm q})}{\omega-E^+_{\bm k} -E^+_{-\bm k+\bm q}+i\delta}-\bar{B}_{\bm k+\bm q}^{\mu\nu}\bar{A}_{\bm k}^{\rho\lambda}\frac{1-f(E^+_{\bm k})-f(E^+_{-\bm k-\bm q})}{\omega+E^+_{\bm k} +E^+_{-\bm k-\bm q}+i\delta}\right], \label{L044}
\end{eqnarray}
where $\hat{\bar{A}}=\hat{\sigma}_3 \hat A$ and
$\hat{\bar{B}}=\hat{\sigma}_3 \hat B$.
In deriving Eq.~(\ref{L044}), we used $E^{-}_{-\bm k}=-E^{+}_{\bm k}$.
We assume $E^{+}_{\bm k}>0$
because we are interested in the stability of Fermi gases before
the pair breaking sets in, i.e., $|\bm v|<|\bm v_{\rm pb}|$.
At $T=0$, the first and second terms in Eq.~(\ref{L044}) vanish from this
condition. The density response functions in Eqs.~(\ref{chiGRPA}) and
(\ref{chiladder}) are calculated by using Eqs.~(\ref{lbar}) and (\ref{L044}).

From Eq.~(\ref{L044}), the spectrum of the particle-hole excitation is given by
\begin{eqnarray}
\omega^{\rm ph}_{\bm q}(\bm k)=E^+_{\bm k} +E^+_{-\bm k+\bm q}. \label{sp}
\end{eqnarray}
For fixed $\bm q$, $\omega^{\rm ph}_{\bm q}(\bm k)$ makes a continuum
for different $\bm k$, as shown in Fig.~\ref{1D_s}. The upper and lower
boundaries of the particle-hole continuum are given by
${\rm min}_{\bm k}\left[\omega^{\rm ph}_{\bm q}(\bm k)\right]$ and ${\rm
max}_{\bm k}\left[\omega^{\rm ph}_{\bm q}(\bm k)\right]$, respectively.

\section{Results}
\label{result}

In this section, by calculating the excitation spectra of the
AB mode and the single-particle excitation, we
discuss the stability of superfluid flow, and determine the critical
velocities of superfluid Fermi gases in 1D, 2D, and 3D optical lattices. 
For this purpose, we calculate the dynamic structure factor
$S_{\bm q}(\omega)=-{\rm Im}[\chi_{\bm q}(\omega)]/\pi$,
which describes the response of the system to density perturbations
with momentum $\bm q$ and frequency $\omega$.
The dynamic structure factor can be directly measured
in experiments by using Bragg spectroscopy \cite{Stenger,Veeravalli}.

Since the GRPA used in this paper is based on a mean-field approximation, it is more 
reliable for higher dimensions.
Nevertheless, calculations of the excitation spectra in the simplest situation of 1D can be useful
for understanding the essence of the physics governing the critical velocity of superfluid fermions
in a lattice.
Hence, we first discuss the excitation spectra and the stability of superfluid flow in 1D lattices.
We note that mean-field theories have been widely used to qualitatively describe excitations 
of trapped atomic gases even in 1D because the finite size ($\sim 100 d$) specific to cold atom 
systems excludes long-wavelength phase fluctuations that destroy the long-range superfluid 
order~\cite{comment2,kraemer,stoeferle}.

\begin{figure}
\centerline{\includegraphics{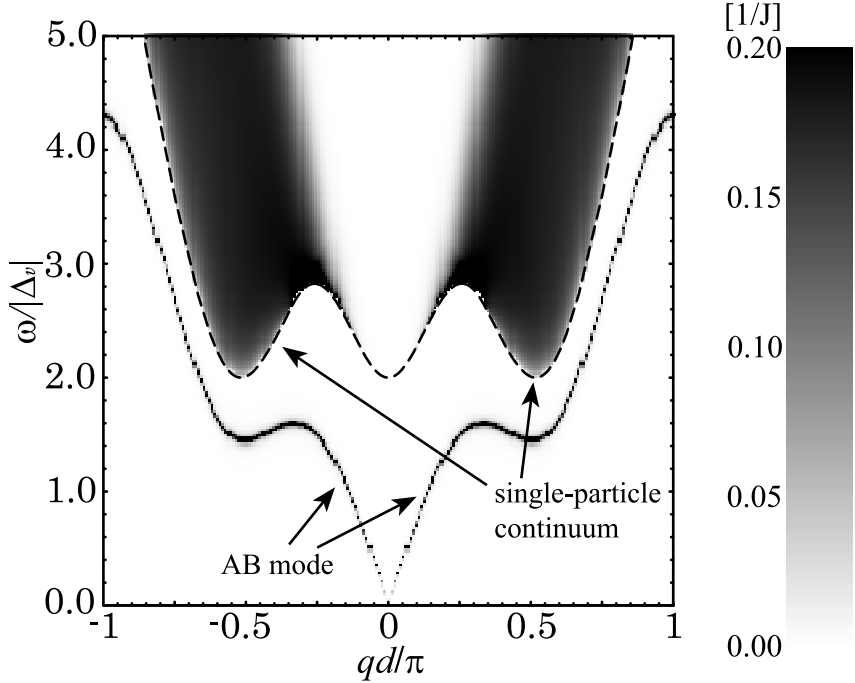}}
\caption{\label{1D_S}Dynamic structure factor $S_q(\omega)$ in 1D
 optical lattices.
The upper gray region and the lower curve correspond to the
particle-hole continuum and the AB mode spectrum, respectively.
The dashed line is the lower boundary of the single-particle
continuum. We set $n=0.5$ (quarter filling), $U=-2.0J$, and $v=0$. 
The superfluid gap and chemical potential are $|\Delta_{v}|=0.409J$ and $\mu=0.624J$.
Here, in numerical calculation, $\delta$ is set to be small 
but finite ($1.0\times 10^{-4}$) so that the peak of the AB mode spectrum has a small finite width.}
\label{1D_s}
\end{figure}

In Fig.~\ref{1D_s}, we show the dynamic structure factor $S_{q}(\omega)$ in 1D optical lattices to illustrate the basic properties of
the excitation spectra.
One clearly sees that the AB mode spectrum lies below the
particle-hole continuum. In addition, the AB mode spectrum has a
characteristic structure with local minima at short wavelengths which
is similar to the roton spectrum in superfluid $^4$He \cite{Griffin}.
Then, it is expected that as the superfluid velocity increases, the energy
of one of the rotonlike minima decreases and it
reaches zero before the lower boundary of the particle-hole continuum does.
This indicates that the instability may be induced by the rotonlike
excitations of the AB mode
rather than by the single-particle excitations because the single-particle
excitations start to have negative energies when the particle-hole
continuum reaches zero energy \cite{Combescot}.
Indeed, we will show that this is the case in all of 1D, 2D, and 3D
optical lattices in the remainder of this section.

We note that this AB mode-induced instability does not occur
in superfluid Fermi gases in uniform 3D systems.
It was found that the instability of superfluid Fermi gases in uniform
3D systems is induced by pair breaking
\cite{Miller,Combescot} which is associated with
the appearance of single-particle excitations with negative energies.
In this case, the AB mode spectrum merges into the particle-hole
continuum in contrast with the behavior in Fig.~\ref{1D_s} where the AB
mode spectrum is separated from the particle-hole continuum.
As a result, the particle-hole continuum reaches zero energy
before the phonon part of the AB mode spectrum starts to have negative
energy as the superfluid velocity increases.
This leads to the instability induced by single-particle excitations,
i.e., by pair breaking. The rotonlike structure of the AB mode spectrum
also appears in a uniform 1D system~\cite{Alm} and a 2D lattice system~\cite{Sofo,Micnas}.

In the following, we first calculate the pair-breaking velocity $v_{\rm
pb}$.
We next discuss the behavior of the AB mode spectrum in 1D, 2D, and 3D
optical lattices by calculating the dynamic structure factor $S_{\bm q}(\omega)$.
To discuss the excitation spectra, we only show the lower boundary of
the particle-hole continuum and the peak of the AB mode spectrum in $S_{\bm
q}(\omega)$ because other details are not
necessary for determining the critical velocities.

\subsection{Pair-breaking velocity}

In this section, we calculate the pair-breaking velocity ${\bm v}_{\rm pb}$, 
which can be analytically obtained from the condition that the
lower boundary of the particle-hole continuum reaches zero energy, i.e.,
${\rm min}_{\bm k}[\omega^{\rm ph}_{\bm q}(\bm k)]=0$~\cite{Combescot}. 

In 1D case, the pair-breaking velocity is given by~\cite{Rodriguez}
\begin{eqnarray}
v_{\rm pb}=\frac{1}{md}\sin^{-1}\left( \frac{|\Delta_v|}{\sqrt{\mu(4J-\mu)}}\right). \label{vpb1D}
\end{eqnarray}
In 2D case, when the supercurrent is flowing in the $(\pi,\pi)$ and
$(\pi,0)$ directions, the pair-breaking velocities are calculated as 
\begin{eqnarray}
|{\bm v}_{\rm pb}|=\frac{\sqrt{2}}{md}\sin^{-1}\left( \frac{|\Delta_{\bm v}|}{\sqrt{\mu(8J-\mu)}}\right) , \label{vpb2D_pipi}
\end{eqnarray}
and
\begin{eqnarray}
|{\bm v}_{\rm pb}|=\left\{ \begin{array}{ll}
{\displaystyle \frac{1}{md}\sin^{-1}\left( \frac{|\Delta_{\bm v}|}{\sqrt{\mu(4J-\mu)}}\right)}, & (\mu<2J), \\
{\displaystyle \frac{1}{md}\sin^{-1}\left( \frac{|\Delta_{\bm v}|}{2J} \right)}, & (\mu\geq 2J), \\
\end{array} \right. \label{vpb2D_pi0}
\end{eqnarray}
respectively. We address the stability of superfluid states in these two cases in Sec.~\ref{2c}.
In 3D case, when the supercurrent is flowing in the $(\pi,\pi,\pi)$ direction, the pair-breaking velocities are calculated as
\begin{eqnarray}
|{\bm v}_{\rm pb}|=\frac{\sqrt{3}}{md}\sin^{-1}\left( \frac{|\Delta_{\bm v}|}{\sqrt{\mu(12J-\mu)}}\right). \label{vpb3D_pipipi}
\end{eqnarray}

Since the order parameter and chemical potential depend on the
superfluid velocity $\bm v$, we must determine $v_{\rm pb}$ by solving
Eq.~(\ref{vpb1D}), (\ref{vpb2D_pipi}), (\ref{vpb2D_pi0}), or (\ref{vpb3D_pipipi}) self-consistently with Eqs.~(\ref{numbereq}) and (\ref{gapeq}). 
In the BCS limit ($|U|\ll J$), $v_{\rm pb}$ approaches zero because
$|\Delta_{\bm v}|$ becomes small.
On the other hand, if $|U|$ is so large that $|\Delta_{\bm v}|$ is
larger than the denominator in $\sin^{-1}$ in Eqs.~(\ref{vpb1D})-(\ref{vpb3D_pipipi}),
$v_{\rm pb}$ is not definable.

\subsection{Stability in 1D optical lattices}
\label{2b}

\begin{figure}
\centerline{\includegraphics{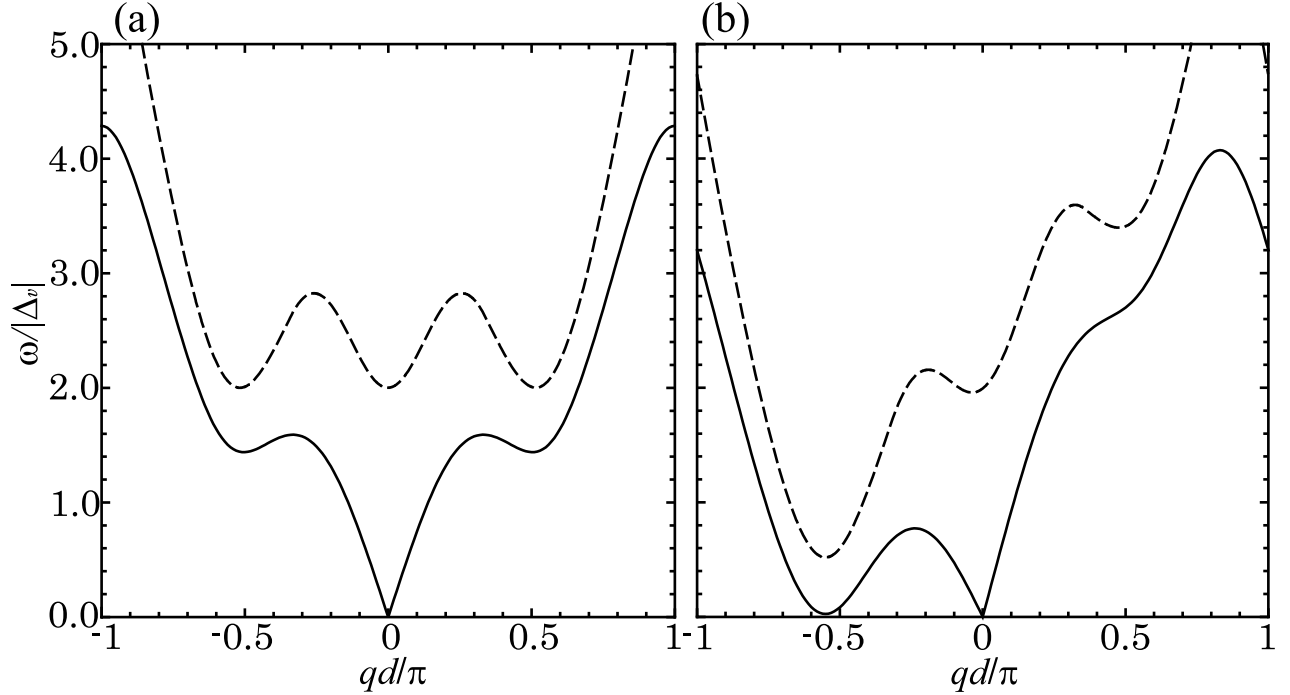}}
\caption{Excitation spectra in 1D optical lattices for (a) current-free
 ($v=0$) and (b) current-carrying ($v=0.21/md$) cases. Solid line
 and dashed line represent the spectrum of the AB mode which
 corresponds to the $\delta$-function peak of $S_{q}(\omega)$ and
 the lower boundary of the particle-hole 
 continuum, respectively. We set $n=0.5$ (quarter filling)
 and $U=-2.0J$. The superfluid gap and chemical potential are 
 (a) $|\Delta_v|=0.409J$ and $\mu=0.624J$, and (b)
 $|\Delta_v|=0.420J$ and $\mu = 0.655J$.  
}
\label{U20n05}
\end{figure}

In this section, we study the stability of superfluid Fermi gases in
1D optical lattices.
In Fig.~\ref{U20n05}, we show the dynamic structure factor $S_q(\omega)$
in 1D optical lattices.
It is clearly seen in Fig.~\ref{U20n05} that the AB mode has a gapless
and linear dispersion in the long-wavelength limit ($|qd|\ll 1$), which
is consistent with the fact that the AB mode is a Nambu-Goldstone mode \cite{Nambu}. 
We obtain the analytic form of the phonon-like dispersion relation in the
long-wavelength limit ($|qd|\ll 1$), as
\begin{eqnarray}
\omega_q&=&d(2J-\mu)\tan(mvd)q \nonumber \\
&&+\frac{1}{\alpha}\sqrt{(1+\alpha)\{v^2_{\rm F}\alpha^2-|\Delta_{v}|^2[\alpha^2+(1+\alpha)\tan^2(mvd)]\}} |q|, \label{ssmv}
\end{eqnarray}
where $v_{\rm F}=2Jd\sin(k_{\rm F}d)\cos(mvd)$, $k_{\rm
F}=|\cos^{-1}\{(2J-\mu)[2J\cos(mvd)]^{-1}\}|/d$, $\alpha=N_{0}U$, and
$N_0=d(\pi v_{\rm F})^{-1}$. 
The details of the derivation of Eq.~(\ref{ssmv}) are summarized in the
Appendix.
When $v=0$, Eq.~(\ref{ssmv}) in the BCS limit ($|\Delta_v| \ll \mu$) 
reduces to the well-known form $\omega_q = v_{\rm F}\sqrt{1+N_{0}U}|q|$,
which was first obtained by Anderson for a uniform 3D system
\cite{Anderson1,Belkhir}
(the dispersion has the coefficient $1/\sqrt{3}$ in 3D case).
We note that Eq.~(\ref{ssmv}) is different from the dispersion relation
obtained in Ref.~\cite{Pitaevskii2} by using the hydrodynamic and
tight-binding approximations.

As we pointed out earlier, in Figs.~\ref{U20n05}(a) and \ref{U20n05}(b),
it is clearly seen that the excitation spectrum of the AB mode lies below the particle-hole continuum, and the AB mode spectrum has roton-like 
minima at $|q|\simeq 2k_{\rm F}^{0}$ [$k_{\rm F}^{0}\equiv n\pi/(2d)$ is
the Fermi wave number in a non-interacting 1D system].
As $v$ increases, the whole spectrum leans toward the left side and the
energy of the rotonlike minimum with $q<0$ decreases as
shown in Fig.~\ref{U20n05}(b). 
As a result, at a certain velocity $v_{\rm c}$, the rotonlike minimum reaches zero energy,
but this occurs before the lower boundary of the particle-hole continuum does.
According to the Landau criterion \cite{Landau}, this indicates that the
spontaneous emission of roton-like excitations of the AB mode is
induced when $v\geq v_{\rm c}$ before pair-breaking occurs at
$v_{\rm pb}$.
Thus, the critical velocity is given by $v_c$ at which the superfluid
flow is destabilized due to the spontaneous emission of rotonlike
excitations of the AB mode.
We note that the phonon part of the AB mode spectrum becomes
negative at a certain velocity larger than $v_{\rm pb}$.
This means that phonon excitations of the AB mode are
irrelevant to the critical velocities.

This instability driven by negative-energy rotonlike excitations of the AB mode corresponds to the energetic instability called Landau instability.
Another type of instability of superfluid called dynamical
instability was also proposed for Fermi gases in optical lattices \cite{Pitaevskii2,Burkov}. 
Dynamical instability is associated with the appearance of complex
energy excitations which was first observed in Bose condensates in
optical lattices \cite{Fallani}.
One can distinguish the dynamical instability from the Landau instability by identifying the 
quasimomentum ${\bm q}$ of the excitations causing the instability.
If the instability is caused by the excitations at the boundary of the first Brillouin zone, e.g., 
$q_x=\pm \pi/d$ or $q_y=\pm\pi/d$ in 2D, it is the dynamical instability because these excitations 
inevitably couple with their anti-phonon branches~\cite{wu,taylor}.
Indeed, we will see that in 2D lattices the dynamical instability due to the AB mode at
${\bm q}=(\pm \pi/d,\pm \pi/d)$ can occur near the half filling or in the BEC region.

\begin{figure}
\centerline{\includegraphics{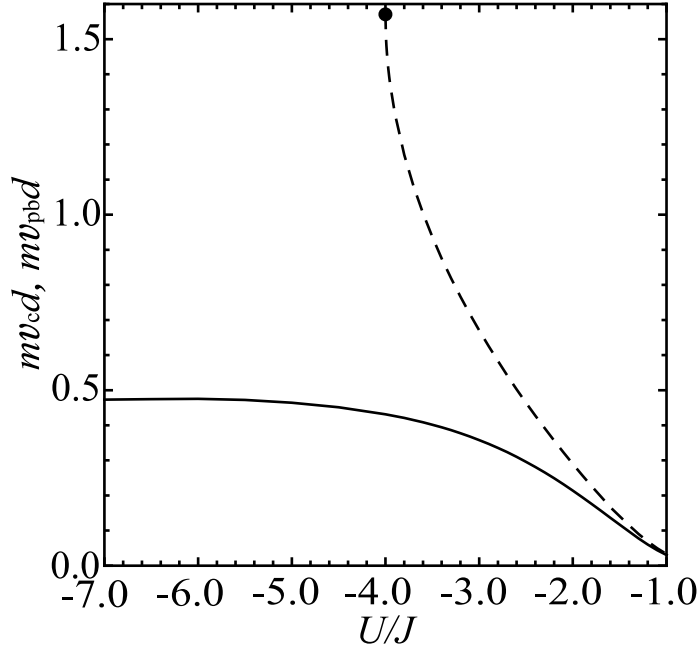}}
\caption{Critical velocity $v_{\rm c}$ (solid line) and pair-breaking
 velocity $v_{\rm pb}$ (dashed line) as functions of $U/J$ in 1D lattices.
 We set $n=0.5$ (quarter filling).
From Eq.~(\ref{vpb1D}), $v_{\rm pb}$ approaches $\pi/2md$ when $|U|/J\rightarrow 4$ (filled circle).
When $|U|/J>4$, the pair-breaking velocity is not definable.
}
\label{vcn05_1D}
\end{figure}

In Fig.~\ref{vcn05_1D}, we show $v_{\rm c}$ and $v_{\rm pb}$ as
functions of $U/J$ in the BCS-BEC crossover region.
One clearly sees that $v_{\rm c}$ is smaller than $v_{\rm pb}$.
We confirmed that $v_{\rm c}$ is smaller than $v_{\rm pb}$ in the entire
BCS region ($-1\lesssim U/J< 0$).
Thus, the instability is always induced by the roton-like excitations in
these regions.
The difference between $v_{\rm c}$ and $v_{\rm pb}$
increases with increasing $|U|/J$ when one approaches the BEC regime.
In addition, both $v_{\rm  c}$ and $v_{\rm pb}$ grow monotonically
with increasing the interaction $|U|/J$.

When $|U|/J \gg 1$ (BEC region), the size of the Cooper pairs becomes smaller than the lattice
spacing and each Cooper pair forms a tightly-bound molecular boson.
In this region, the Hubbard model of Eq.~(\ref{Hubbard}) is mapped onto a hardcore 
Bose-Hubbard model with nearest-neighbor repulsive interactions (or equivalently the spin-$\frac{1}{2}$ {\it XXZ} model)~\cite{Alexandrov,Anderson1,Burkov,Sofo}.
Since it is well known that any kinds of mean-field theory completely fail to describe hardcore bosons
in 1D, our GRPA is also invalid  in the BEC limit in 1D.
Hence, we postpone the discussion of the BEC limit to the next section, where we will show results
in 2D.

\begin{figure}
\centerline{\includegraphics{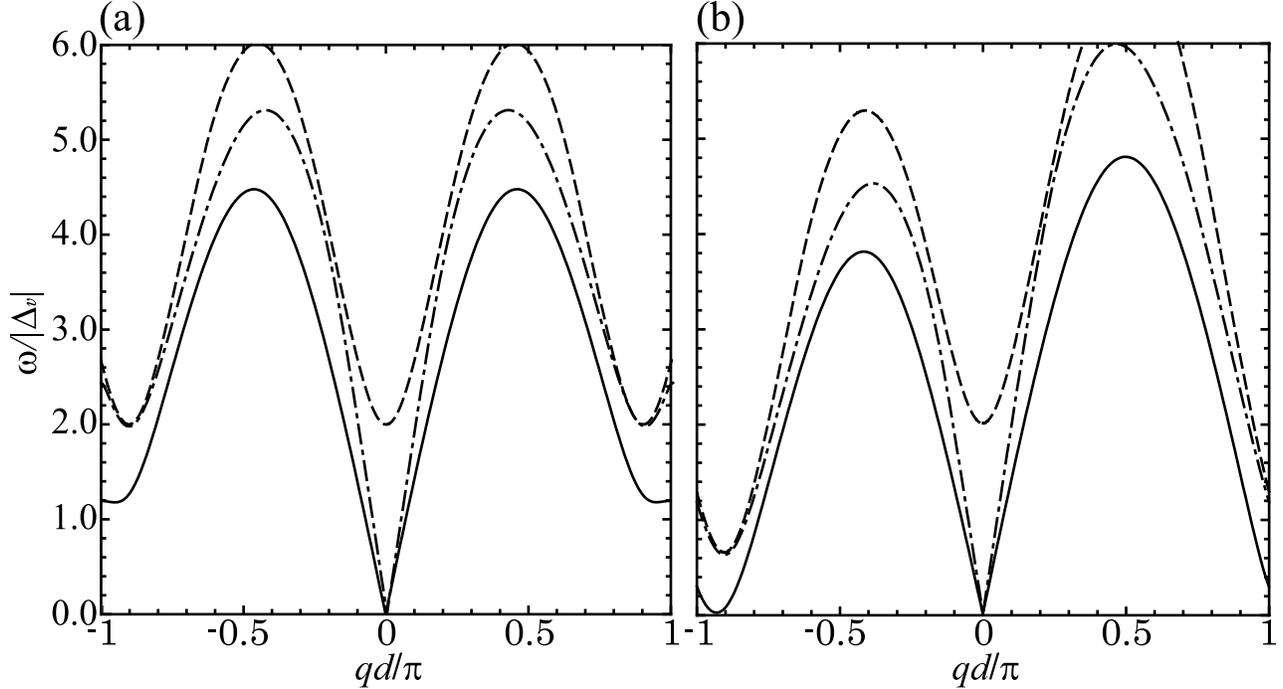}}
\caption{Excitation spectra in 1D optical lattices for (a) current-free
 ($v=0$) and (b) current-carrying ($v=0.12/md$) cases. Solid line,
 dash-dotted line, and dashed line represent the spectrum of the
 AB mode obtained from Eq.~(\ref{chiGRPA}), Eq.~(\ref{chiladder}), and the lower boundary of the single-particle
 excitation continuum, respectively. We set $n=0.9$ and $U=-2.0J$. The superfluid gap and chemical potential are calculated
 as (a) $|\Delta_{v}|=0.345J$ and $\mu=1.69J$, and (b)
 $|\Delta_{v}|=0.350J$ and $\mu = 1.70J$.  
}
\label{U20n09}
\end{figure}

\begin{figure}
\centerline{\includegraphics{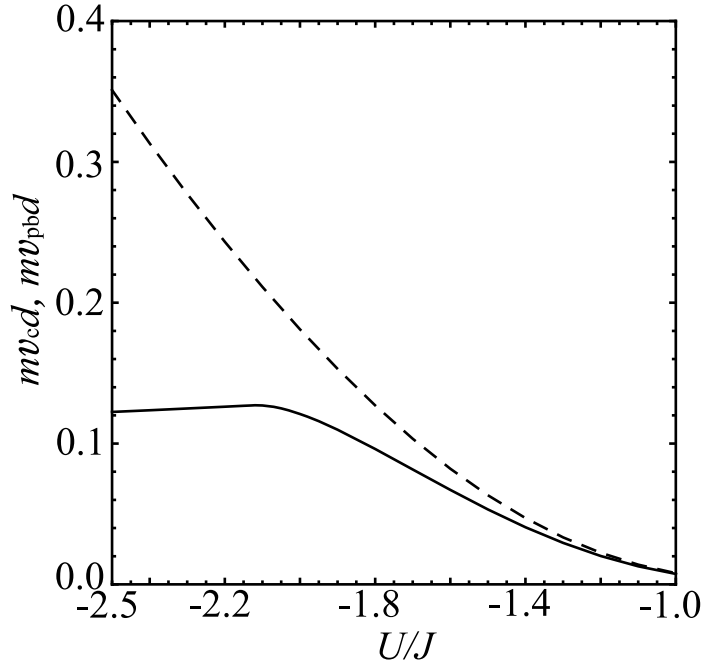}}
\caption{Critical velocity $v_{\rm c}$ (solid line) and pair-breaking
 velocity $v_{\rm pb}$ (dashed line) as functions of $U/J$ in 1D
 lattices near half filling. We set $n=0.9$.
}
\label{vcn09_1D}
\end{figure}

To discuss the origin of the roton-like minima of the AB spectrum, we
show the dynamic structure factor when $n=0.9$ in Fig.~\ref{U20n09}.
Compared to the AB mode spectrum when $n=0.5$ in Fig.~\ref{U20n05}, the roton-like
minima have lower energies than those in Fig.~\ref{U20n05}.
It turns out that as one approaches half filling ($n=1$), the energy of the
rotonlike minima becomes smaller.
As is well known, the fluctuation due to the formation of
charge-density-wave (CDW) order is enhanced near half filling in
lattice fermion systems \cite{Giamarchi}.
Thus, the CDW fluctuation leads to the rotonlike minima in the AB mode spectrum.
At half filling, the rotonlike minima reach zero energy even in the
current-free case ($v=0$) and the superfluid ground state becomes unstable
due to the formation of CDW order.
We show the critical velocity $v_c$ when $n=0.9$ as a function of $U/J$
in Fig.~\ref{vcn09_1D}.
The critical velocity when $n=0.9$ in Fig.~\ref{vcn09_1D} is smaller than
that when $n=0.5$ in Fig.~\ref{vcn05_1D} due to the strong CDW
fluctuation. We notice that $v_c$ becomes almost constant below a
certain value of interaction ($U/J\lesssim -2$), which reflects the fact
that the energy difference between the lower boundary of the
particle-hole continuum and the roton-like minimum becomes large as $|U|$ increases. 
This indicates that the CDW fluctuation is enhanced below this value of
interaction. 

To support the above consideration on the origin of the roton-like minimum,
we compare the AB mode spectra calculated by Eqs.~(\ref{chiGRPA}) and~(\ref{chiladder}) in Fig.~\ref{U20n09}.
As discussed in Sec.~\ref{responsefn}, Eq.~(\ref{chiladder}) includes only
the contribution from the ladder diagrams, while Eq.~(\ref{chiGRPA})
includes the contributions both from ladder and RPA-type bubble
diagrams.
In Fig.~\ref{U20n09}, one clearly sees that the AB mode spectrum
calculated by Eq.~(\ref{chiGRPA}) lies below the one calculated by
Eq.~(\ref{chiladder}) which actually lies close to the lower boundary of the
particle-hole continuum. 
Since the RPA-type bubble diagrams include the effect of the CDW fluctuation
\cite{Giamarchi}, this behavior of the AB mode spectrum is consistent
with the above consideration that the CDW fluctuation leads to the
roton-like structure of the AB mode spectrum.

\subsection{Stability in 2D optical lattices}
\label{2c}
\begin{figure}
\centerline{\includegraphics{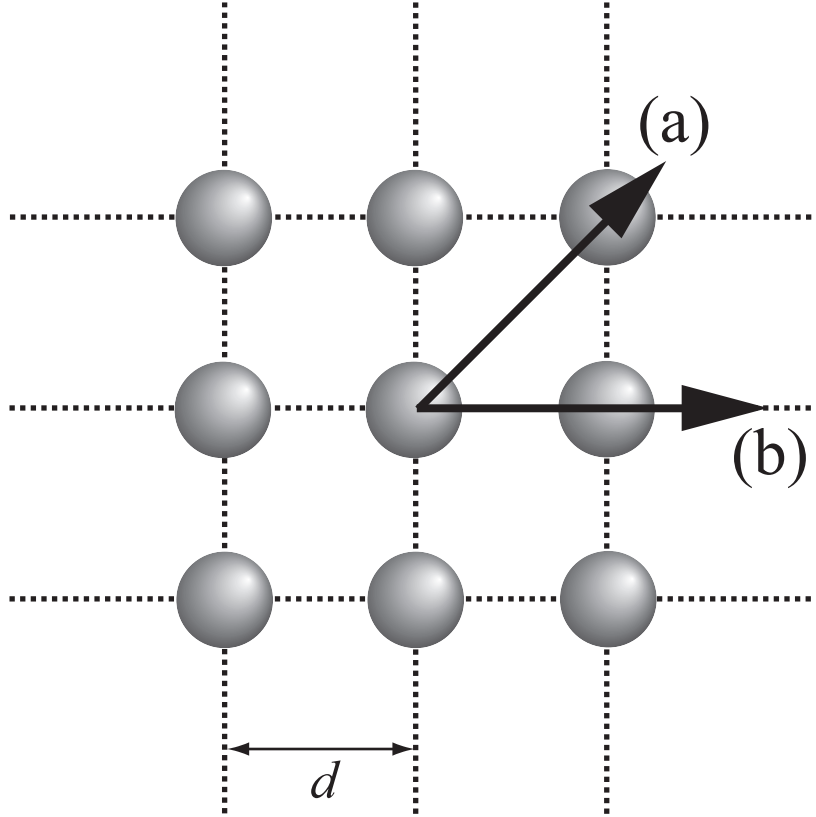}}
\caption{Schematic picture of 2D optical lattices. Arrows indicate
 the superfluid velocity in (a) $(\pi,\pi)$ and (b) $(\pi,0)$ directions.
 Each circle represents a lattice site.} 
\label{moving}
\end{figure}

In this section, we discuss the stability of Fermi gases in 2D
optical lattices. 
Here, we restrict ourselves to two characteristic situations where the superfluid flows along the $(\pi,\pi)$ or $(\pi,0)$ directions (see Fig.~\ref{moving}) in order to see the effects of CDW fluctuations on the stability of superfluid Fermi gases.

First, we discuss the case when the superfluid flows along the
$(\pi,\pi)$ direction. We assume the superfluid velocity $\bm
v=(v,v)/\sqrt{2}$, where $v\equiv |{\bm v}|$. We calculate the dynamic
structure factor $S_{\bm q}(\omega)$ only when $q_x=q_y$ because 
superfluid flow is expected to be most unstable for excitations with momenta in the opposite direction to the flow.

\begin{figure}
\centerline{\includegraphics{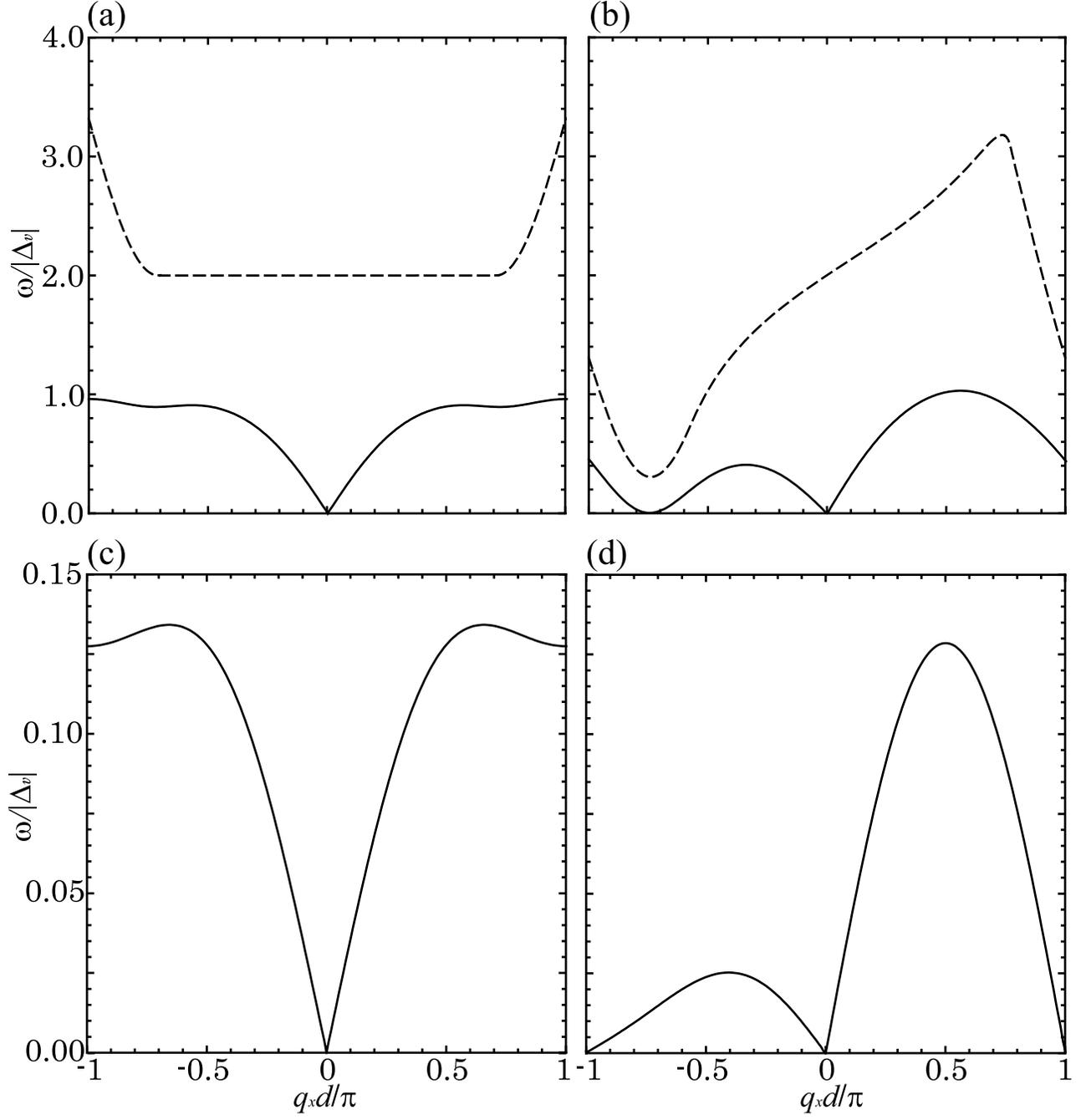}}
\caption{Excitation spectra in 2D optical
 lattices when the superfluid flows along the $(\pi,\pi)$ direction.
 Solid and dashed lines represent the
 spectrum of the AB mode and the lower boundary of the particle-hole 
 continuum, respectively.
 The superfluid velocity is [(a),(c)] $v=0$, (b) $v=0.467/md$, and
 (d) $v=0.6502/md$. We set $n=0.5$, $q_x=q_y$, [(a),(b)]$U=-4.5J$, and [(c),(d)]$U=-12.0J$. 
 The superfluid gap and chemical potential are calculated as 
 (a) $|\Delta_{\bm v}|=1.33J$ and $\mu=2.23J$, (b) $|\Delta_{\bm v}|=1.39J$ and $\mu
 =2.29J$, (c) $|\Delta_{\bm v}|=4.92J$ and $\mu=0.687J$, and (d) $|\Delta_{\bm v}|=4.97J$ and $\mu=0.746J$.  
}
\label{pipi}
\end{figure}

In Figs.~\ref{pipi}(a) and \ref{pipi}(b), we show the excitation spectra when $n=0.5$ and $q_x=q_y$.
When $v=0$ [see Fig.~\ref{pipi}(a)], the rotonlike structure is slightly seen
in the AB mode spectrum. 
As $v$ increases, the rotonlike structure becomes remarkable and one of
the rotonlike minima goes down. At a certain velocity $v_c$ smaller
than the pair-breaking velocity $v_{\rm pb}$, the energy
of the rotonlike minimum reaches zero [see Fig.~\ref{pipi}(b)].
Thus, as 1D case, the critical velocity is given by $v_c$ at which the
instability due to spontaneous emission of rotonlike excitations of the
AB mode sets in.

In Figs~\ref{pipi}(c) and \ref{pipi}(d), we show the excitation spectra in the BEC region 
($U=-12J$)~\cite{comment3}.
There we see that the rotonlike minima of the AB mode are present also in this region.
 The critical velocity in the BEC region is also determined by the rotonlike excitations
[see Fig.~\ref{pipi}(d)].
Since the roton-like minima are shifted to $q_x=q_y=\pm \pi/d$, the instability caused by the 
roton-like excitations is the dynamical instability~\cite{comment1}.
The shift of the roton-like minima to the edge of the Brillouin zone can be understood as follows.
As mentioned before, in the BEC region the Hubbard model can be reduced to a hardcore 
Bose-Hubbard model with nearest-neighbor repulsive interactions.
The nearest-neighbor repulsion enhances density wave fluctuations with the wave vector 
${\bm k}=(\pi/d,\pi/d)$~\cite{scalettar}, leading to the rotonlike minimum at ${\bm q}=(\pi/d,\pi/d)$.
Thus, the shift of the roton minima means that as one approaches the BEC region, the origin of
the roton minima changes from the nesting effect of the Fermi surface to the nearest-neighbor
interactions between molecular bosons.
Notice that in the limit of the low filling ($n\rightarrow0$), the roton minima of the AB mode do not 
survive any longer~\cite{Sofo} and the critical velocity is determined by the long-wavelength part 
(phonon branch) of the AB mode.

\begin{figure}
\centerline{\includegraphics{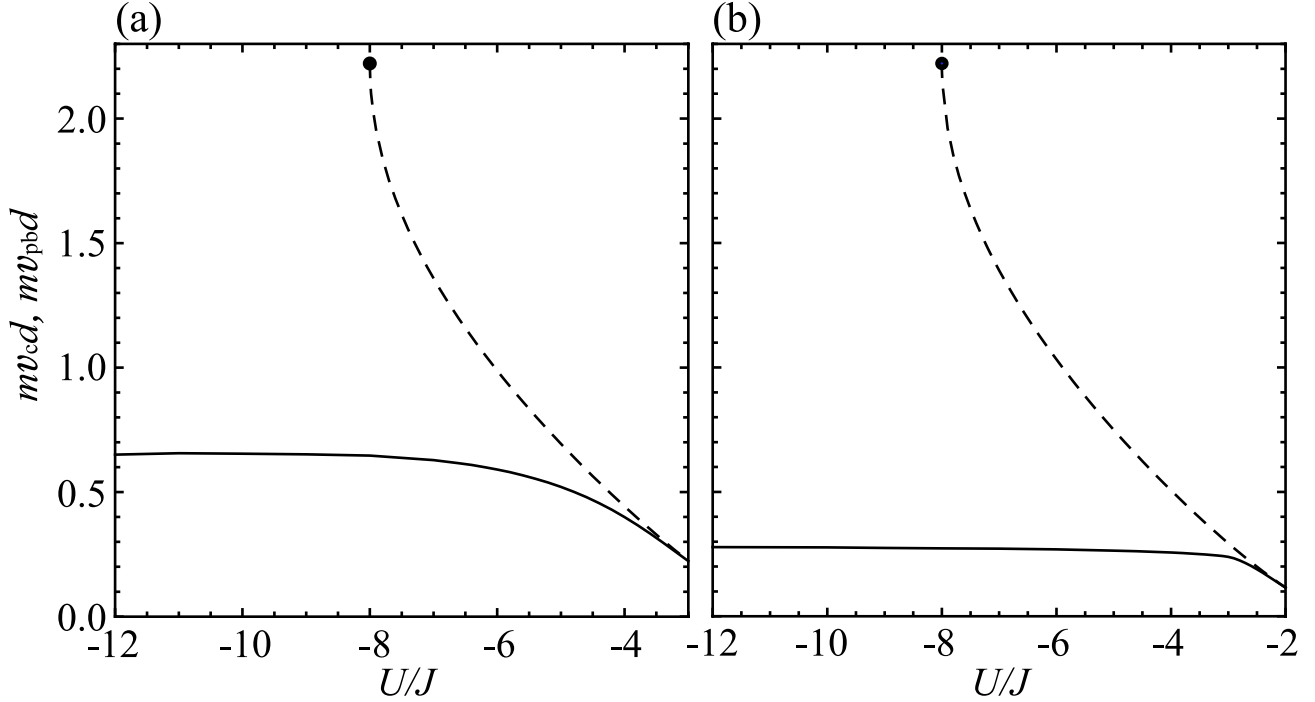}}
\caption{Critical velocity $v_{\rm c}$ (solid line) and pair-breaking
 velocity $v_{\rm pb}$ (dashed line) as functions of $U/J$ in 2D
 lattices when the superfluid flows along the $(\pi,\pi)$ direction. We set
 (a) $n=0.5$ and (b) $n=0.8$.
From Eq.~(\ref{vpb2D_pipi}), $v_{\rm pb}$ approaches $\sqrt{2}\pi/2md$ when $|U|/J\rightarrow 8$ (filled circle).
When $|U|/J>8$, the pair-breaking velocity is not definable.
}
\label{vcpipi}
\end{figure}
In Fig.~\ref{vcpipi}, we show the critical velocity $v_c$ and the pair-breaking
velocity $v_{\rm pb}$ as functions of $U/J$. $v_c$ and $v_{\rm pb}$ show
qualitatively the same behavior as in 1D case. Namely, $v_c$ is smaller
than $v_{\rm pb}$ and they increase monotonically with increasing the
interaction strength $|U|/J$. 

\begin{figure}
\centerline{\includegraphics{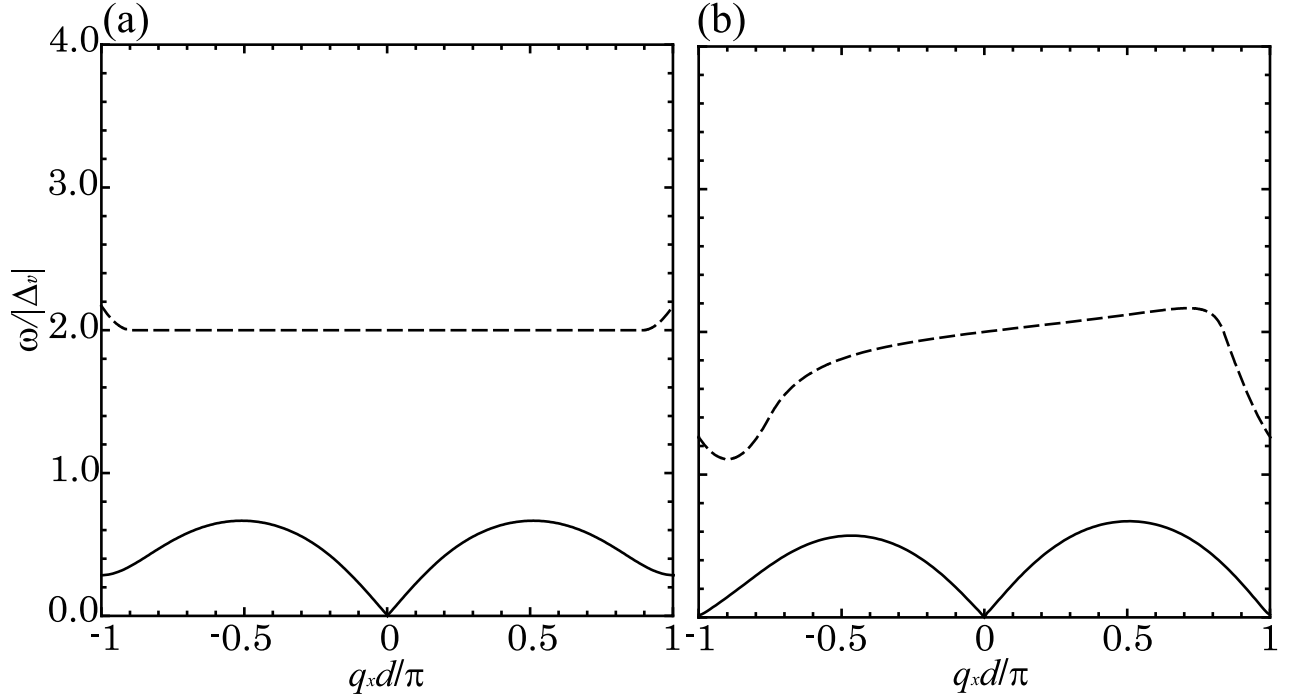}}
\caption{Excitation spectra in 2D optical
 lattices when the superfluid flows along the $(\pi,\pi)$ direction.
 Solid and dashed lines represent the
 spectrum of the AB mode and the lower boundary of the particle-hole 
 continuum, respectively.
 The superfluid velocities are (a) $v=0$ and (b) $v=0.260/md$.
The filling is $n=0.8$. 
 We set $U=-4.5J$ and $q_x=q_y$. 
 The superfluid gap and chemical potential are calculated as 
 (a) $|\Delta_{\bm v}|=1.60J$ and $\mu=3.32J$, and (b) $|\Delta_{\bm v}|=1.62J$ and $\mu=3.33J$.  
}
\label{pipihigh}
\end{figure}

Near half filling, the rotonlike structure when $v=0$ becomes more remarkable 
due to strong CDW fluctuation [see Fig.\ref{pipihigh}(a)].
Since it is seen in Fig.~\ref{pipihigh}(b) that the instability is driven by the rotonlike excitations with 
${\bm q}=(\pm \pi/d, \pm \pi/d)$, it is the dynamical instability.
As a result of the strong CDW fluctuation, the critical velocity near half filling is smaller
than the one at quarter filling, as shown in Fig.~\ref{vcpipi}(b).
As in 1D case, the CDW fluctuation is strongly enhanced below a certain value of interaction ($U/J\lesssim -3$) in Fig.~\ref{vcpipi}(b).
As a result, $v_c$ is almost constant below this value of interaction.

\begin{figure}
\centerline{\includegraphics{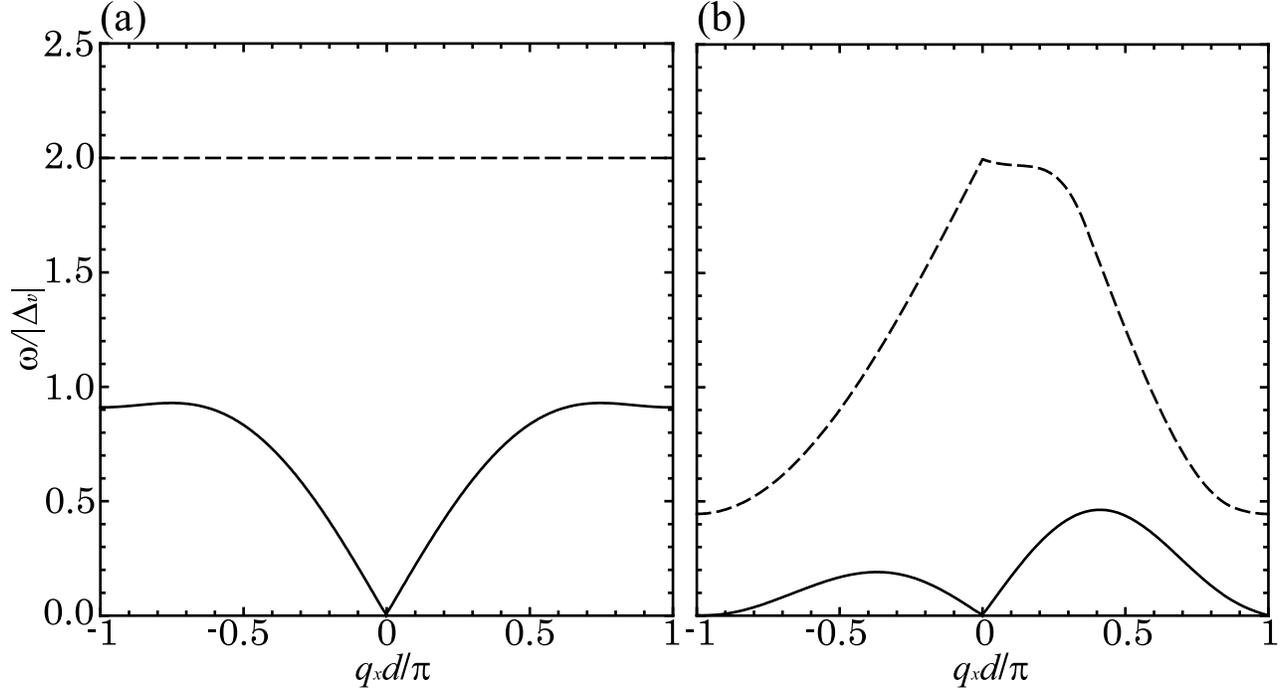}}
\caption{Excitation spectra in 2D optical
 lattices when the superfluid flows along the $(\pi,0)$ direction
 for (a) current-free ($v=0$) and (b) current-carrying ($v=0.585/md$) cases.
 Solid and dashed lines represent the
 spectrum of the AB mode and the lower boundary of the particle-hole 
 continuum, respectively.
 We set $n=0.5$, $U=-4.5J$, and $q_y=0$. 
 The superfluid gap and chemical potential are calculated as
 (a) $|\Delta_{\bm v}|=1.33J$ and $\mu=2.23J$, and (b) $|\Delta_{\bm v}|=1.41J$ and $\mu=2.31J$.  
}
\label{pi0}
\end{figure}

Next, we discuss the case when the supercurrent flows along the
$(\pi,0)$ direction in 2D optical lattices (see Fig.~\ref{moving}).
We assume the superfluid velocity as $\bm v=(v,0)$.
The excitation spectra when $n=0.5$ is shown in
Fig.~\ref{pi0}. Here, we assume $q_y=0$ from the same reason for the $(\pi,\pi)$ case.
The behavior of the energy spectra is qualitatively the same as the previous cases,
i.e., the AB mode spectrum has the rotonlike structure and lies below the
particle-hole continuum. As the superfluid velocity increases, the AB mode spectrum is pushed down and the energy of the rotonlike minimum decreases.
The instability sets in at $v_c$ when the energy of the roton-like minimum becomes zero.

\begin{figure}
\centerline{\includegraphics{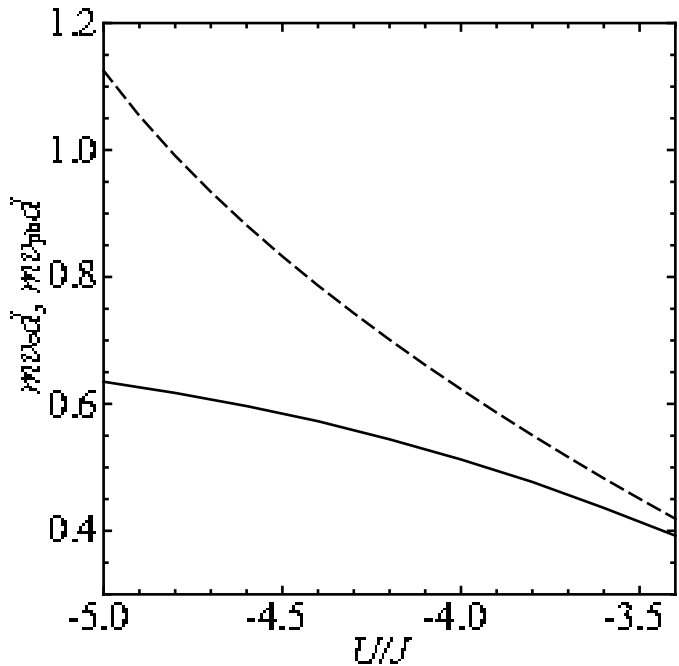}}
\caption{Critical velocity $v_{\rm c}$ (solid line) and pair-breaking
 velocity $v_{\rm pb}$ (dashed line) as functions of $U/J$ in 2D
 lattices when the superfluid flows along the $(\pi,0)$ direction.
 We set $n=0.5$.
}
\label{vcpi0}
\end{figure}

In Fig.~\ref{vcpi0}, we show the critical velocity $v_c$ and pair-breaking
velocity $v_{\rm pb}$ as functions of $U/J$. They also show
qualitatively the same behavior as the previous cases.
Comparing Figs.~\ref{vcpipi} and \ref{vcpi0}, we find that $v_c$ in the
$(\pi,\pi)$ case is smaller than the one in the $(\pi,0)$ case.
This is because the nesting effect of the Fermi surface occurs in
the $(\pi,\pi)$ direction so that it is enhanced by the supercurrent
in parallel to this direction.

We note that near the half filling, the AB mode spectrum with $q_y\neq0$ [i.e., not in the $(\pi,0)$ direction] may reach zero
before that in the $(\pi,0)$ direction does, even when
the superfluid flows in the $(\pi,0)$ direction. This is also due to the strong nesting effect in the $(\pi,\pi)$ direction. However, even in this case, our main conclusion remains unchanged.  Namely, the instability is induced by the AB mode excitations. This effect was also pointed out in the BEC regime~\cite{Burkov}.

\subsection{Stability in 3D optical lattices}
\label{2d}
\begin{figure}
\centerline{\includegraphics{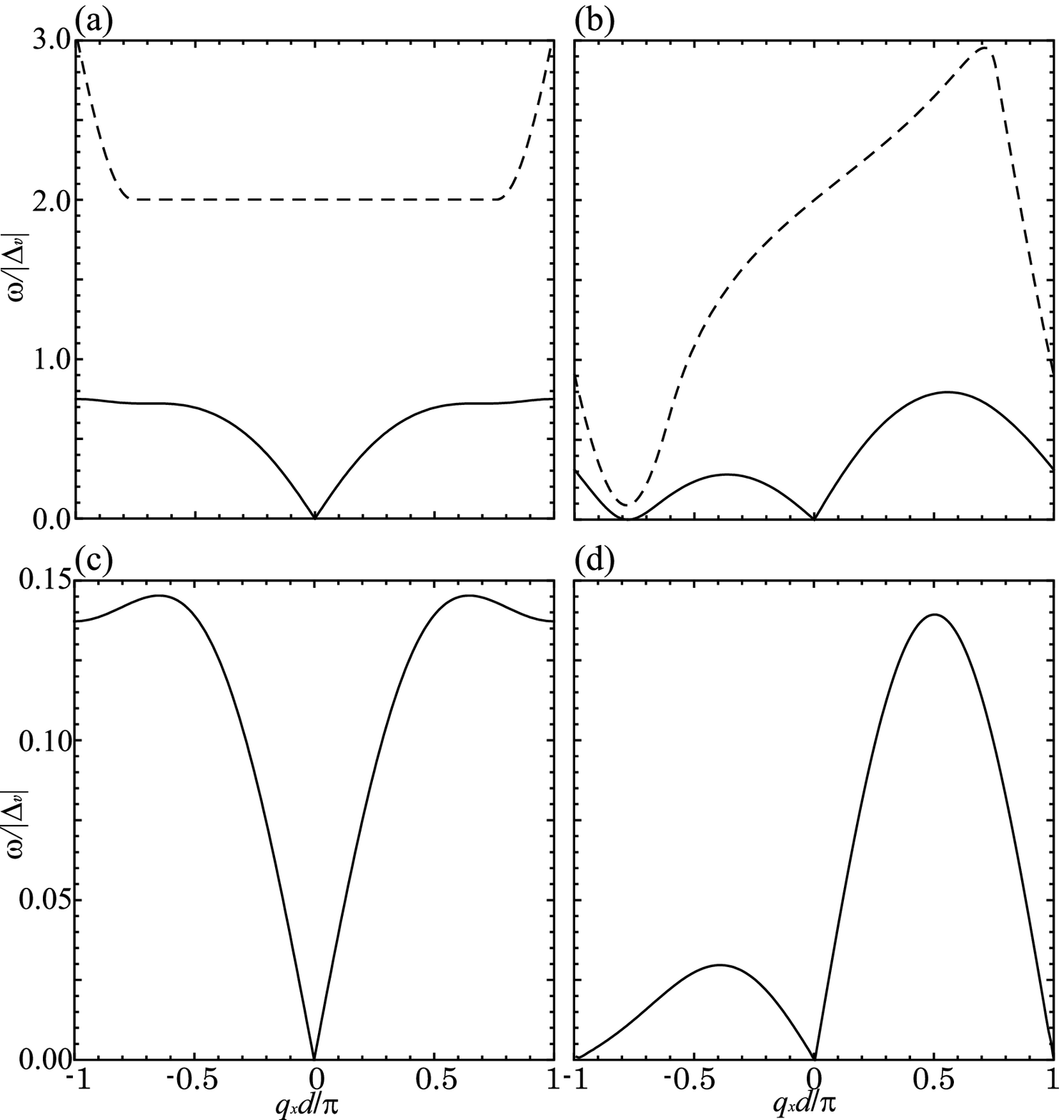}}
\caption{Excitation spectra in 3D optical
 lattices when the superfluid flows along the $(\pi,\pi,\pi)$ direction
 for [(a),(c)] $v=0$, (b) $v=0.5969/md$, and (d) $v=0.7894/md$.
 Solid and dashed lines represent the
 spectrum of the AB mode and the lower boundary of the particle-hole 
 continuum, respectively.
 We set $n=0.5$, $q_x=q_y=q_z$, [(a),(b)] $U=-6.0J$, and [(c),(d)] $U=-14.0J$. 
 The superfluid gap and chemical potential are calculated as
 (a) $|\Delta_{\bm v}|=1.92J$ and $\mu=3.78J$, (b) $|\Delta_{\bm v}|=1.98J$ and $\mu=3.85J$, (c) $|\Delta_{\bm v}|=5.70J$ and $\mu=2.11J$, and (d) $|\Delta_{\bm v}|=5.77J$ and $\mu=2.18J$.
}
\label{3Dpipipi}
\end{figure}

\begin{figure}
\centerline{\includegraphics{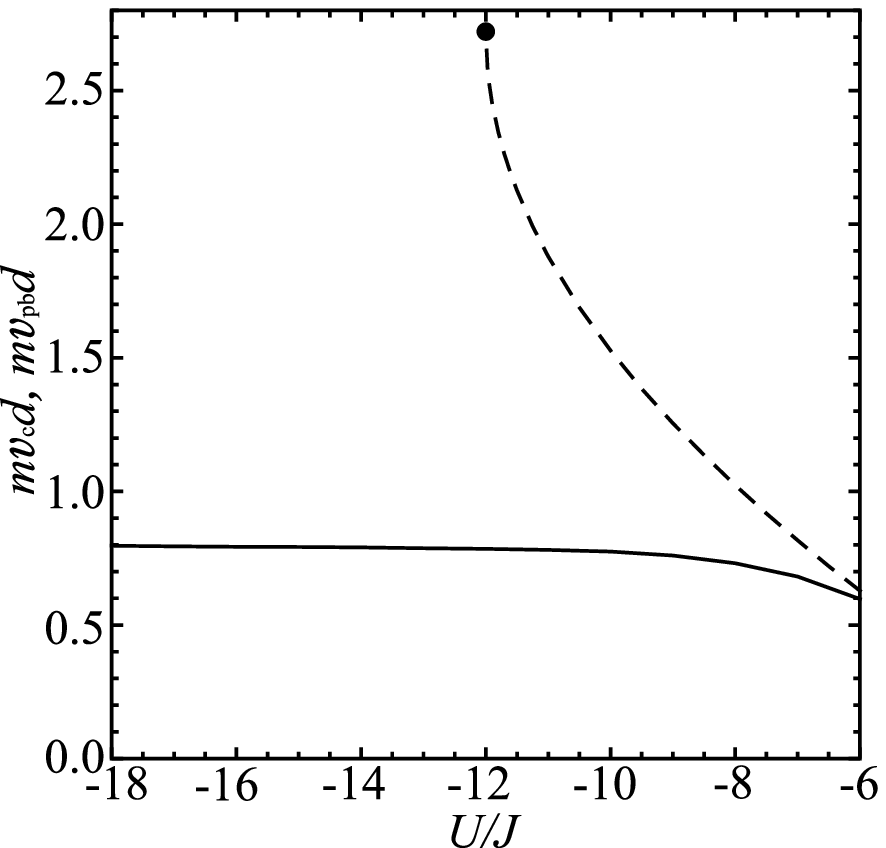}}
\caption{Critical velocity $v_{\rm c}$ (solid line) and pair-breaking
 velocity $v_{\rm pb}$ (dashed line) as functions of $U/J$ in 3D
 lattices when the superfluid flows along the $(\pi,\pi,\pi)$ direction.
 We set $n=0.5$.
From Eq.~(\ref{vpb3D_pipipi}), $v_{\rm pb}$ approaches $\sqrt{3}\pi/2md$ when $|U|/J\rightarrow 12$ (filled circle).
When $|U|/J>12$, the pair-breaking velocity is not definable.
}
\label{vcpipipi}
\end{figure}

Let us finally discuss the stability of Fermi gases in 3D optical lattices.
We restrict ourselves to the situation where the superfluid flows along the $(\pi,\pi,\pi)$ direction. 
We show the excitation spectra in Fig.~\ref{3Dpipipi} for $q_x=q_y=q_z$ 
and the critical velocity and the pair-breaking velocity as functions of $U/J$ in Fig.~\ref{vcpipipi}.
It is clearly seen that the AB mode lies well below the single-particle continuum and that in the 
entire region of the attractive interaction, the roton part of the AB mode reaches zero before the 
single-particle continuum does.
This leads to the conclusion that the critical velocity of superfluid Fermi gases in a deep lattice is 
determined by the roton part of the AB mode except in the low filling limit, regardless of the dimensionality of the system.

\section{conclusion}
\label{conclusion}

In conclusion, we have studied the stability of superfluid Fermi gases
in 1D, 2D, and 3D optical lattices at $T=0$. By applying the GRPA Green's
function formalism developed by C\^ot\'e and Griffin \cite{Cote} to the
attractive Hubbard model, we
calculated the excitation spectra of the AB mode as well as the single-particle excitation in the
presence of superfluid flow. 
We found that the AB mode spectrum has the characteristic rotonlike
structure being separated from the particle-hole continuum due to the
strong CDW fluctuation. The energy of the rotonlike
minimum decreases as the superfluid velocity increases and it reaches
zero at the critical velocity before pair-breaking occurs. 
This indicates that the instability of superfluid flow in 1D, 2D, and 3D
optical lattices is induced by the spontaneous emission of the rotonlike excitations of the AB mode.
We calculated the critical velocity $v_c$ as functions of $U/J$ and
confirmed that it is smaller than the pair-breaking velocity $v_{\rm
pb}$ in the BCS and BCS-BEC crossover region.
We also found that the CDW instability is strongly enhanced near
half filling which leads to the suppression of the critical velocity
when the attractive interaction is large.

Finally, we remark that our results are valid for superfluid Fermi gases in
{\it deep} optical lattices because we employed the tight-binding Hubbard
model. From this reason, our results for 1D optical lattices cannot be directly compared to the experiment with shallow optical lattices in Ref.~\cite{Miller}.
However, superfluid Fermi gases have been already achieved in deep
optical lattices in Ref.~\cite{Chin}.
Our theoretical predictions in this paper may be verified if a superfluid
Fermi gas is prepared in a moving deep optical lattice.

\acknowledgments

We wish to thank K. Kamide, K. Osawa, M. Takahashi, T. Kimura, T. Nikuni, and S. Kurihara for useful discussions and comments.
We acknowledge Y. Ohashi and M. Tsubota for valuable comments.
I.D. and D.Y. are supported by a Grant-in-Aid from JSPS.

\begin{appendix}
\section{AB phonon spectrum in 1D}
\label{AB1D}

Here, we give a detailed derivation of the AB mode spectrum in 1D in the long-wavelength limit in Eq.~(\ref{ssmv}).
The matrix $\hat{\mathcal D}_q(i\Omega_n\to\omega)$ in Eq.~(\ref{matrixD}) is given as
\begin{eqnarray}
\hat{\mathcal D}_{\bm q}(\omega)=\left(
\begin{array}{cccc}
a+b & d+e & f-g & c \\
-d-e & -h-i & -c & -f-g \\
-f+g & -c & i-h & -d+e \\
c & f+g & d-e & a-b \\
\end{array} 
\right),
\end{eqnarray}
where
\begin{eqnarray}
a&=&\frac{1}{M}\sum_{k}\frac{\mathcal{EE'}-\bar{\xi}'\bar{\xi}}{2\mathcal{EE'}} \frac{\mathcal{E+E'}}{(\omega-\eta'+\eta)^2-(\mathcal{E+E'})^2},\label{a}\\
b&=&\frac{1}{M}\sum_{k}\frac{\mathcal{E}\bar{\xi}'-\mathcal{E}'\bar{\xi}}{2\mathcal{EE'}} \frac{\omega-\eta'+\eta}{(\omega-\eta'+\eta)^2-(\mathcal{E+E'})^2},\label{b}\\
c&=&\frac{1}{M}\sum_{k}\frac{\Delta_{v}^2}{2\mathcal{EE'}} \frac{\mathcal{E+E'}}{(\omega-\eta'+\eta)^2-(\mathcal{E+E'})^2},\\
d&=&\frac{1}{M}\sum_{k}\frac{\Delta_{v}}{2\mathcal{E}} \frac{\omega-\eta'+\eta}{(\omega-\eta'+\eta)^2-(\mathcal{E+E'})^2},\\
e&=&\frac{1}{M}\sum_{k}\frac{\Delta_{v}\bar{\xi}'}{2\mathcal{EE'}} \frac{\mathcal{E+E'}}{(\omega-\eta'+\eta)^2-(\mathcal{E+E'})^2},\\
f&=&\frac{1}{M}\sum_{k}\frac{\Delta_{v}}{2\mathcal{E'}} \frac{\omega-\eta'+\eta}{(\omega-\eta'+\eta)^2-(\mathcal{E+E'})^2},\\
g&=&\frac{1}{M}\sum_{k}\frac{\Delta_{v}\bar{\xi}}{2\mathcal{EE'}} \frac{\mathcal{E+E'}}{(\omega-\eta'+\eta)^2-(\mathcal{E+E'})^2},\\
h&=&\frac{1}{M}\sum_{k}\frac{\mathcal{EE'}+\bar{\xi}'\bar{\xi}}{2\mathcal{EE'}} \frac{\mathcal{E+E'}}{(\omega-\eta'+\eta)^2-(\mathcal{E+E'})^2},\\
i&=&\frac{1}{M}\sum_{k}\frac{\mathcal{E}\bar{\xi}'+\mathcal{E}'\bar{\xi}}{2\mathcal{EE'}}
 \frac{\omega-\eta'+\eta}{(\omega-\eta'+\eta)^2-(\mathcal{E+E'})^2}.\label{i}
\label{element}
\end{eqnarray}
Here, we used the notations $\mathcal{E}\equiv\mathcal{E}_{k}$,
$\mathcal{E}'\equiv\mathcal{E}_{k+q}$, 
$\bar{\xi}\equiv\bar{\xi}_{k}$, $\bar{\xi}'\equiv\bar{\xi}_{k+q}$,
$\eta\equiv\eta_{k}$, and $\eta'\equiv\eta_{k+q}$. 
For simplicity, we took $\Delta_{v}$ to be real.
We calculate the matrix elements Eq. (\ref{a})-(\ref{i}) in the limit of $qd\ll 1$ and $\omega/J \ll 1$.
Expanding Eq. (\ref{a}) to the second order in $q$ and $\omega$, $a$ is
obtained as
\begin{eqnarray}
a&\simeq&\frac{\Delta_{v}^2}{4M}\sum_{k}\left[-\frac{1}{\mathcal{E}^3}+qd \frac{3J\bar{\xi}\sin(kd)\cos(mvd)}{\mathcal{E}^5}-\omega^2\frac{1}{4\mathcal{E}^5} +qd\omega\frac{J\cos(kd)\sin(mvd)}{\mathcal{E}^5}\right.\nonumber\\
&&\left.+(qd)^2\left( \frac{J^2\Delta_{v}^2\sin^2(kd)\cos^2(mvd)}{\mathcal{E}^7}-\frac{J^2\cos^2(kd)\sin^2(mvd)}{\mathcal{E}^5}+\frac{3J\bar{\xi}\cos(kd)\cos(mvd)}{2\mathcal{E}^5} \right) \right].
\end{eqnarray}
By carrying out the integration over $k$, we obtain
\begin{eqnarray}
a&\simeq&
 -\frac{N_0}{2}-\frac{N_0\omega^2}{12\Delta_{v}^2}+qd\omega\frac{N_0J\cos(k_{\rm
 F} d)\sin(mvd)}{3\Delta_{v}^2}\nonumber\\
&&-(qd)^2N_0\frac{(3\Delta_{v}^2-16J\mu+4\mu^2)\cos(2mvd)}{48\Delta_{v}^2 \cos^2(mvd)}
-(qd)^2N_0\frac{\Delta_{v}^2-16J^2\sin^4(mvd)}{48\Delta_{v}^2 \cos^2(mvd)}.\label{a2}
\end{eqnarray}
Here, we have used the standard approximation,
$\frac{1}{2\pi}\int^{\pi/d}_{-\pi/d} dk\ F(k)\rightarrow
N_0\int^{\infty}_{-\infty}d\bar{\xi}\ F(\bar{\xi})$, where $F(k)$ is an
arbitrary function. 
By calculating Eqs. (\ref{b})-(\ref{i}) in the same way as Eq. (\ref{a}), we obtain
\begin{eqnarray}
b&\simeq& 0, \label{b2}\\
c&\simeq& -\frac{N_0}{2}-\frac{N_0\omega^2}{12\Delta_{v}^2}+qd\omega\frac{N_0J\cos(k_{\rm F} d)\sin(mvd)}{3\Delta_{v}^2} \nonumber\\
&&+(qd)^2\frac{2J(4\mu-J)+8J^2\sin^4(mvd)}{24\Delta_{v}^2}-(qd)^2\frac{(16J^2+\Delta_{v}^2)\sin^2(mvd)}{24\Delta_{v}^2}\\
d&\simeq& f\simeq -\frac{N_0}{4\Delta_{v}}[ \omega-2qdJ\cos(k_{\rm F} d)\sin(mvd) ],\\
e&\simeq& -qd\omega\frac{\tan(mvd)}{12\Delta_{v}}+(qd)^2\frac{(2J-\mu)[3+2\tan^2(mv)]}{24\Delta_{v}},\\
g&\simeq& -qd\omega\frac{\tan(mvd)}{12\Delta_{v}}-(qd)^2\frac{(2J-\mu)[3-2\tan^2(mv)]}{24\Delta_{v}},\\
h&\simeq& \frac{1}{U}+\frac{N_0}{2}-\frac{N_0\omega^2}{6\Delta_{v}^2}+qd\omega\frac{2N_0J\cos(k_{\rm F} d)\sin(mvd)}{3\Delta_{v}^2} \nonumber\\
&&+(qd)^2N_0\frac{13\Delta_{v}^2\sin^2(mvd)-16J^2\sin^4(mvd)}{24\Delta_{v}^2\cos^2(mvd)}\nonumber\\
&&-(qd)^2N_0\frac{3\Delta_{v}^2+2J^2-8J\mu+2\mu^2}{12\Delta_{v}^2\cos^2(mvd)},\\
i&\simeq& \frac{qd\tan(mvd)}{2}\left(N_{0}+\frac{1}{U} \right).\label{i2}
\end{eqnarray}
Note that when we calculate $1/M\sum_k 1/\mathcal{E}$, we have used the
gap equation [Eq. (\ref{gapeq})] to eliminate the divergence.
Assuming that the AB mode has a linear dispersion relation in the
long-wavelength limit ($qd\ll 1$) and calculating the pole of 
Eq.~(\ref{chiGRPA}) by using Eqs. (\ref{a2})-(\ref{i2}), we obtain Eq.~(\ref{ssmv}).

\end{appendix}


\begin{thebibliography}{}
\bibitem{Regal} 
C. A. Regal, M. Greiner, and D. S. Jin, Phys. Rev. Lett. \textbf{92}, 040403 (2004).

\bibitem{Bartenstein}
M. Bartenstein, A. Altmeyer, S. Riedl, S. Jochim, C. Chin, J. H. Denschlag, and R. Grimm, Phys. Rev. Lett. \textbf{92}, 120401 (2004).

\bibitem{Zwierlein}
M. W. Zwierlein, C. A. Stan, C. H. Schunck, S. M. F. Raupach,
	A. J. Kerman, and W. Ketterle, Phys. Rev. Lett. \textbf{92}, 120403 (2004).

\bibitem{Kinast}
J. Kinast, S. L. Hemmer, M. E. Gehm, A. Turlapov, and J. E. Thomas,
	Phys. Rev. Lett. \textbf{92}, 150402 (2004).

\bibitem{Bourdel}
T. Bourdel, L. Khaykovich, J. Cubizolles, J. Zhang, F. Chevy,
	M. Teichmann, L. Tarruell, S. J. J. M. F. Kokkelmans, and
	C. Salomon, Phys. Rev. Lett. \textbf{93}, 050401 (2004).

\bibitem{CChin}
C. Chin, M. Bartenstein, A. Altmeyer, S. Riedl, S. Jochim, J. Hecker
	Denschlag, R. Grimm, Science \textbf{305}, 1128 (2004).

\bibitem{Giorgini}
S. Giorgini, L. Pitaevskii, and S. Stringari, Rev. Mod. Phys. \textbf{80}, 1215 (2008).

\bibitem{Ketterle}
W. Ketterle and M. W. Zwierlein,
Proceedings of the International School of Physics ``Enrico Fermi,''
	Course CLXIV, edited by M. Inguscio, W. Ketterle, and C. Salomon
	(IOS Press, Amsterdam, 2008).


\bibitem{Eagles}
D. M. Eagles, Phys. Rev. \textbf{186}, 456 (1969).

\bibitem{Leggett} 
A. J. Leggett, {\it Modern Trends in the Theory of Condensed Matter} (Springer, Berlin, 1980).

\bibitem{Nozieres}
P. Nozi\`eres and S. Schmitt-Rink, J. Low Temp. Phys. \textbf{59}, 195 (1985).

\bibitem{SadeMelo}
C. A. R. Sa de Melo, M. Randeria, and J. R. Engelbrecht, Phys. Rev. Lett. \textbf{71}, 3202 (1993).

\bibitem{Holland}
M. Holland, S. J. J. M. F. Kokkelmans, M. L. Chiofalo, and R. Walser, Phys. Rev. Lett. \textbf{87}, 120406 (2001).

\bibitem{Timmermans}
E. Timmermans, K. Furuya, P. W. Milonni, and A. K. Kerman, Phys. Lett. A \textbf{285}, 228 (2001).

\bibitem{Ohashi}
Y. Ohashi and A. Griffin, Phys. Rev. Lett {\textbf 89}, 130402 (2002); Phys. Rev. A \textbf{67}, 033603
	(2003); Phys. Rev. A \textbf{67}, 063612 (2003).

\bibitem{Tamaki}
H. Tamaki, Y.Ohashi, and K. Miyake, Phys. Rev. A \textbf{77}, 063616 (2008).

\bibitem{Pitaevskii1}L. P. Pitaevskii and S. Stringari, {\it
	Bose-Einstein Condensation} (Oxford Science Publications, Oxford, 2003).
\bibitem{Tilley} 
D. R. Tilley and J. Tilley, {\it Superfluidity and Superconductivity} (Hilger, Bristol, 1991).

\bibitem{Vollhardt} 
D. Vollhardt and P. W\"olfle, {\it The Superfluid Phases of Helium 3}, (Taylor \& Francis, London, 1990).

\bibitem{Raman}
C. Raman, M. K\"ohl, R. Onofrio, D. S. Durfee, C. E. Kuklewicz,
	Z. Hadzibabic, and W. Ketterle, Phys. Rev. Lett. \textbf{83},
	2502 (1999).

\bibitem{Onofrio}
R. Onofrio, C. Raman, J. M. Vogels, J. R. Abo-Shaeer, A. P. Chikkatur,
	and W. Ketterle, Phys. Rev. Lett. \textbf{85}, 2228 (2000).


\bibitem{Tinkham}
M. Tinkham, {\it Introduction to Superconductivity}, (McGraw-Hill, New York, 1975).

\bibitem{Landau}
L. D. Landau,
J. Phys. (USSR) \textbf{5}, 71 (1941).

\bibitem{Miller}
D. E. Miller, J. K. Chin, C. A. Stan, Y. Liu, W. Setiawan, C. Sanner, and W. Ketterle,
Phys. Rev. Lett. \textbf{99}, 070402 (2007).

\bibitem{Combescot}
R. Combescot, M. Yu. Kagan, and S. Stringari,
Phys. Rev. A \textbf{74}, 042717 (2006).

\bibitem{Spuntarelli}
A. Spuntarelli, P. Pieri, and G. C. Strinati, Phys. Rev. Lett. 
{\textbf 99}, 040401 (2007).

\bibitem{Pitaevskii2} 
L. P. Pitaevskii, S. Stringari, and G. Orso
Phys. Rev. A \textbf{71}, 053602 (2005).

\bibitem{Cote}
R. C\^ot\'e and A. Griffin,
Phys. Rev. B \textbf{48}, 10404 (1993).

\bibitem{Anderson1} 
P. W. Anderson, 
Phys. Rev. \textbf{112}, 1900 (1958).

\bibitem{Bogoliubov} 
N. N. Bogolyubov, 
Sov. Phys. Usp. \textbf{2}, 236 (1959).

\bibitem{Belkhir} 
L. Belkhir and M. Randeria, 
Phys. Rev. B \textbf{49}, 6829 (1994).

\bibitem{Koponen} 
T. Koponen, J.-P. Martikainen, J. Kinnunen and P. T\"orm\"a, 
Phys. Rev. A \textbf{73}, 033620 (2006).

\bibitem{KadanoffBaym}
L. P. Kadanoff and G. Baym, {\it Quantum Statistical Mechanics}
	(W. A. Benjamin, New York, 1962).

\bibitem{Gorkov}
L. P. Gorkov, Sov. Phys. JETP \textbf{7}, 505 (1958).

\bibitem{Nambu}
Y. Nambu, Phys. Rev. \textbf{117}, 648 (1960).

\bibitem{Rodriguez}
M. Rodriguez and P. T\"orm\"a,
Phys. Rev. A \textbf{69}, 041602 (2004).

\bibitem{Stenger}
J. Stenger, S. Inouye, A. P. Chikkatur, D. M. Stamper-Kurn, D. E. Pritchard, and W. Ketterle,
Phys. Rev. Lett. \textbf{82}, 4569 (1999).

\bibitem{Veeravalli}G. Veeravalli, E. Kuhnle, P. Dyke, and C. J. Vale,
	Phys. Rev. Lett. \textbf{101}, 250403 (2008).
	
\bibitem{comment2}
For example, the Gross-Pitaevskii mean-field theory qualitatively explained an experiment of excitation
creation in 1D superfluid Bose gases in a vibrating optical lattice~\cite{kraemer,stoeferle} except for the hardcore boson regime.

\bibitem{kraemer}
M. Kr\"amer, C. Tozzo, and F. Dalfovo, 
Phys. Rev. A {\bf 71}, 061602(R) (2005).

\bibitem{stoeferle}
T. St\"oferle, H. Moritz, C. Schori, M. K\"ohl, and T. Esslinger,
Phys. Rev. Lett. 92, 130403 (2004).

\bibitem{Griffin}
A. Griffin, {\it Excitations in a Bose-Condensed Liquid} (Cambridge, 1993).

\bibitem{Alm} 
T. Alm and P. Schuck, 
Phys. Rev. B \textbf{54}, 2471 (1996).

\bibitem{Sofo}
J. O. Sofo, C. A. Balseiro, and H. E. Castillo, 
Phys. Rev. B \textbf{45}, 9860 (1992).

\bibitem{Micnas}
T. Kostyrko and  R. Micnas, 
Phys. Rev. B \textbf{46}, 11025 (1992);
Acta Phys. Pol. \textbf{A83}, 837 (1993).

\bibitem{Burkov}
A. A. Burkov and A. Paramekanti,
Phys. Rev. Lett. \textbf{100}, 255301 (2008).

\bibitem{Fallani}
L. Fallani, L. De Sarlo, J. E. Lye, M. Modugno, R. Saers, C. Fort, and M. Inguscio,
Phys. Rev. Lett. \textbf{93}, 140406 (2004).

\bibitem{wu}
B. Wu and Q. Niu, 
Phys. Rev. A {\bf 64}, 061603(R) (2001).

\bibitem{taylor}
E. Taylor and E. Zaremba,
Phys. Rev. A {\bf 68}, 053611 (2003).

\bibitem{Alexandrov}
A. Alexandrov and J. Ranninger, 
Phys. Rev. B \textbf{23}, 1796 (1981).

\bibitem{Giamarchi} 
T. Giamarchi, {\it Quantum Physics in One Dimension} (Oxford University Press, Oxford, 2004).

\bibitem{comment3}
The validity of the GRPA in the BEC region has been confirmed for 2D and 3D cases
by comparing the excitation spectra calculated by GRPA to those obtained by applying
the linear-spin wave approximation to the corresponding spin model~\cite{Alexandrov,Sofo}.

\bibitem{comment1}
Recently, it has been reported that an instability caused by this excitation is 
the dynamical instability also in the following paper: 
R. Ganesh, A. Paramekanti, and A. A. Burkov, Phys. Rev. A {\bf 80} 043612 (2009).

\bibitem{scalettar}
R. T. Scalettar, G. G. Batrouni, A. P. Kampf, and G. T. Zimanyi,
Phys. Rev. B {\bf 51}, 8467 (1995).

\bibitem{Chin}
J. K. Chin, D. E. Miller, Y. Liu, C. Stan, W. Setiawan, C. Sanner, K. Xu, and W. Ketterle, Nature \textbf{443}, 961 (2006).

\end{thebibliography}
\end{document}